\documentclass[reprint, amsmath, amssymb, aps]{revtex4-2}

\usepackage{graphicx}
\usepackage{dcolumn}
\usepackage{bm}
\usepackage[colorlinks,linkcolor=blue,anchorcolor=blue,citecolor=blue]{hyperref}

\begin{document}

\title{Absorption and scattering of a high dimensional non\textendash{}commutative black hole}

\author{Mao\textendash{}Yuan Wan}
\email{maoyuan.wan.physics@gmail.com}
\affiliation{
University of Shanghai for Science and Technology, Shanghai 200093, China
}

\author{Chen Wu}
\email{wuchenoffd@gmail.com}
\affiliation{
Shanghai Advanced Research Institute, Chinese Academy of Sciences, Shanghai 201210, China
}

\date{\today}

\begin{abstract}

In this work, 
we investigate the scattering of massless plane scalar waves by the high dimensional non\textendash{}commutative Schwarzschild\textendash{}Tangherlini black hole. 
We use the partial wave approach to determine the scattering and absorption cross sections in the incident wavelength range. 
Our numerical results demonstrate that the bigger the non\textendash{}commutative parameter, 
the smaller the maximum value of the related partial absorption cross section, 
however the tendency is slightly. 
We also discovered that when the non\textendash{}commutative parameter is weak, 
the absorption cross section of the high dimensional black hole oscillates in the low frequency zone. 
The total absorption cross section oscillates around the geometrical optical limit in the high frequency range, 
and the scattering characteristics of black holes with various parameters are visibly different. 
The influence on the differential scattering cross section is particularly pronounced at large angles.

\end{abstract}

\maketitle

\section{Introduction} \label{sec1}

The problem of understanding how to avoid singularities in black hole spacetimes is important in general relativity. 
Researchers have been very interested in non\textendash{}commuting black holes as an alternative to quantum gravity \cite{bib1,bib2}. 
The theory arose from the perspective of string theory \cite{bib3}. 
The existence of a minimal length is a natural need related to quantum mechanics when considering the quantum properties of the phase space. 
As a result, 
the presence of singularities can be prevented by constraining the minimum scale of space and time, 
i.e. 
replacing the point\textendash{}like matter distribution with a smeared matter distribution \cite{bib4,bib5}. 
Obviously, 
black holes are an essential application area for non\textendash{}commutative geometry, 
and understanding non\textendash{}commutative can help us better comprehend black hole evaporation. 
Susskind \cite{bib6} suggested in 1993 that the string effect should not be neglected in the string and black hole correlation theory. 
String effects in black holes inspired by non\textendash{}commutative geometry are analogous to non\textendash{}commutative field theory derived from string theory in some aspects \cite{bib7}. 
In 1999, 
Seiberg and Witten \cite{bib8} claimed that non\textendash{}commutative field theory can explain some open\textendash{}string low\textendash{}energy efficient theories with nontrivial backgrounds, 
so providing a theoretical foundation for the study of non\textendash{}commutative spacetime. 
String theory studies show that spacetime coordinates become a non\textendash{}commutative operator on the $D$\textendash{}brane. 
Its commutator is
\begin{equation} \label{eq1}
[\hat{x}_{\mu},\hat{x}_{\nu}]=i\theta_{\mu\nu}=i\frac{C_{\mu\nu}}{\varLambda^{2}_{\text{NC}}},
\end{equation}
where $\hat{x}_{\mu}$ is the spacetime coordinate operator and $\theta_{\mu\nu}$ is an antisymmetric constant tensor with dimension equal to the square of the length, 
$\varLambda^{2}_{\text{NC}}$ is the mass scale associated with non\textendash{}commutative and $C^{ab}$, 
and it is commonly thought of as a frame\textendash{}free dimensionless antisymmetric matrix which is not a tensor. 
Eq. (\ref{eq1}) may do the non\textendash{}commutative correction of quantum field theory. 

Nicolini and his colleagues \cite{bib7} point out that non\textendash{}commutativity is a characteristic of the manifold itself, 
not a superposition of geometric forms. 
As a result, 
non\textendash{}commutative affects the source term and has no effect on the Einstein tensor section of the field equation. 
That is, 
the Gaussian smeared matter distribution \cite{bib7} replaces the mass density of the point\textendash{}like function of the energy\textendash{}momentum tensor of the Einstein equations. 
In addition to the Gaussian smeared matter distribution, 
there exist the Lorentz \cite{bib9} and Rayleigh distributions \cite{bib10}. 
This results in a non\textendash{}commutative self\textendash{}regular black hole solution devoid of singularities. 
Nicolini and his colleagues \cite{bib7} became the first to solve a 4\textendash{}dimensional non\textendash{}commutative geometrically inspired Shwarzschild black hole. 
It is then extended to charged \cite{bib11} case and extra spatial dimensions \cite{bib12} case. 
Modesto and Nicolini \cite{bib13} extended it to the general situation of electrically charged rotating non\textendash{}commutative black holes in 2010. 
Non\textendash{}commutative black holes also extend to higher dimensional black holes \cite{bib14,bib15} and higher dimensional charged black holes \cite{bib16,bib17}. 
Nozari and Mehdipour \cite{bib9} investigated the Hawking radiation of non\textendash{}commutative Shwarzschild black holes. 
Many other researchers have investigated the thermodynamic aspects of non\textendash{}commutative black holes \cite{bib18,bib19,bib20,bib21}.
Yan \cite{bib21ad1,bib21ad2} used the WKB method to calculate the QNMs of non\textendash{}commutative black hole.
Konoplya and Bronnikov et al. \cite{bib21ad3,bib21ad4} used WKB method to study Quasinormal ringing of regular black holes.

Since non\textendash{}commutative spacetime coordinates introduce a new basic natural length scale, 
it is also fascinating to investigate the influence of this constant parameter on matter wave absorption and scattering in black hole spacetime. 
During the 1970s and 1980s, 
considerable results of attempt \cite{bib22} was given to the study of the scattering of various planar waves of frequency $\omega$ by black holes. 
People can regard the scattering results as a black hole fingerprint and possibly observed. 
Recently, 
these topics have been also arousing lots of interest \cite{bib23,bib24,bib25,bib26,bib27}.
However, 
there are far too many fields around horizons in real astrophysical complicated black holes. 
In order to understand black holes as much as possible, 
we should consider as many fields as possible, 
but this will complicate the calculation.
We might as well simplify the problem for this.
The simplest instance, 
the scalar wave in the Spherically symmetrical black hole, 
immediately displays several basics of the problem,
such as the high frequency limit of the absorption cross section tends to the area of the black hole shadow or Hawking radiation.
For increasingly complicated cases, 
the analytical intricacy might obscure the calculation's basic simplicity.
Thus,  
we will look at massless scalar waves.
Can the computations of scattering be used in astrophysics? 
Regarding the impure, 
astrophysically complicated black holes, 
we must agree that observation has yet to provide us with a completely convincing candidate to which we might apply these ideas.
Considering that photons are only detect a black hole outside the shadow due to the spacetime structure and the astrophysical environment surrounding it. 
The underlying geometrical structure could only be detected by gravitational wave \cite{bib22}.
In theory, high-precision measurements of massless scalar wave fluxes scattered by black holes might one day be used to estimate black hole's hairs.
Even in the absence of experimental confirmation, 
we believe that research into black hole scattering will continue to further our knowledge of how black holes interact with their surroundings.

The reason that this subject attracts increasing attentions are as follows: 
1)quasinormal modes of black holes are thought to representing the poles of the corresponding black hole scattering matrix. 
2)scattering and absorption from black holes is related to many interesting phenomena, 
such as glory, 
photon orbit and superradiant scattering \cite{bib24,bib28,bib29,bib30,bib31,bib32,bib33,bib34,bib35,bib36,bib37}. 
In Ref. \cite{bib38,bib39,bib40}, 
S\'{a}nchez noticed that the absorption cross section of the Schwarzschild black hole for massless scalar wave oscillating near the geometrical optical limit via numerical approaches.

In recent years, 
the partial wave approach has been widely employed in the research of black hole scattering, particularly the scattering process of numerous fields of black holes \cite{bib41,bib42,bib43,bib44,bib45}. 
Chen and his coworkers \cite{bib46} investigated the scattering and absorption cross section of massless scalar wave from the black hole surrounded by  magnetic field. 
Crispino et al. \cite{bib47} extended study of the scattering theory of scalar waves in the Reissner\textendash{}Nordstr\"{o}m black hole spacetime. 
Huang et al. \cite{bib48} studied the scattering and absorption cross section of massless scalar waves by the Bardeen black hole. 
It is worth noting that the Bardeen black hole is a type of regular black hole, 
as well as one of the non\textendash{}singularity black hole theories. 
Anacleto et al. \cite{bib49} investigates the scattering and absorption cross sections of non\textendash{}commutative Schwarzschild black holes with Gaussian smeared matter distribution and Lorentz smeared matter distribution.
Non\textendash{}commutative BTZ black holes also cause scattering \cite{bib50}. 
Scalar wave scattering by spherically symmetric $D$\textendash{}dimensional black holes is also studied in string theory \cite{bib51}.

The major focus of this study is on the scalar scattering process of the $D$\textendash{}dimensional non\textendash{}commutative Schwarzschild\textendash{}Tangherlini black hole with the smeared matter distribution, 
as well as the influence of different $\theta$ values on the scalar absorption and scattering cross sections.
The following is the structure of this paper: 
Sect. \ref{sec2} presents the perturbation equation and effective potential equation for scalar perturbation in a given spacetime background and briefly discusses the $D$\textendash{}dimensional non\textendash{}commutative Schwarzschild\textendash{}Tangherlini black hole with smeared matter distribution. 
We also explore the allowed range of hoop parameter $k$ under the hoop hypothesis, 
as well as the range of non\textendash{}commutative parameter $\theta$ corresponding to different dimension $D$ and value $k$ when event horizon exist. 
In Sect. \ref{sec3}, 
we perform an analysis of the classical dynamics of massless scalar field. 
The concentration of Sect. \ref{sec3} is the absorption and scattering cross sections of massless scalar fields for the $D$\textendash{}dimensional non\textendash{}commutative Schwarzschild\textendash{}Tangherlini black holes. 
Finally, 
we come to a conclusion in Sect. \ref{sec4}. 
Throughout this paper we use natural units $c=\hbar=G=k_{B}=1$.

\section{The basic equation} \label{sec2}

\subsection{The metric} \label{subsec2.1}

Non\textendash{}commutative geometry is represented physically as a fluid that smeared rather than is squeezed at the origin, 
and the energy\textendash{}momentum tensor changed by the smeared matter distribution corresponds to the anisotropy fluid. 
Many researchers \cite{bib52,bib53,bib54,bib55} have recently demonstrated that the Gaussian smeared matter distribution may be substituted as long as the origin has a pointed peak comparable to the Dirac $\delta$\textendash{}function and the integral of the distribution function is finite. 
The author provide the general density of smeared matters in Ref. \cite{bib55},
\begin{equation} \label{eq2}
\rho_{\text{matter}}(r)=\left[\frac{M}{\pi^{\frac{D-1}{2}}(4\theta)^{\frac{D+k-1}{2}}}\cdot\frac{\Gamma\left(\frac{D-1}{2}\right)}{\Gamma\left(\frac{D+k-1}{2}\right)}\right]r^{k}e^{-\frac{r^{2}}{4\theta}},
\end{equation}
where $M$ is the black hole's mass and $theta$ is a positive non\textendash{}commutative parameter. 
$k$ is a non\textendash{}negative integer, 
and the Gaussian smeared matter distribution corresponds to $k=0$, 
the Rayleigh distribution corresponds to $k=1$, 
the Maxwell\textendash{}Boltzmann distribution corresponds to $k=2$, 
and so on. 
$\Gamma(x)$is the gamma function.

The average radius corresponding to the mass density distribution of the matter can be calculated using Eq. (\ref{eq2}),
\begin{equation} \label{eq3}
\bar{r}=\int\limits_{0}^{+\infty}\frac{\rho_{\text{matter}}(r)}{M}r\mathrm{d}V_{D-1}=\sqrt{4\theta}\frac{\Gamma\left(\frac{D+k}{2}\right)}{\Gamma\left(\frac{D+k-1}{2}\right)},
\end{equation}
where $\mathrm{d}V_{D-1}$ represents the $(D-1)$ dimension volume element. 
The greater the corresponding value $\theta$, 
the more disperse the matter distribution; 
the smaller the corresponding parameter $\theta$, 
the more concentrated the matter distribution. 
The mean radius of matter is $\bar{r}\to0$ in the limit of $\theta\to0$, 
which indicates that the distribution of matter collapses into a point and the non\textendash{}commutative of spacetime disappears. 
As a result, 
this spacetime non\textendash{}commutative can be described as a tiny influence overlaid on ordinary spacetime, 
with the non\textendash{}commutative impact mostly reflected in the area around the average radius of matter.

The following requirements must be met for the geometric metric solution of a static spherically symmetric non\textendash{}commutative Schwarzschild black hole: 
1) The radial matter distribution function $\rho_{\text{matter}}(r)$ is spherically symmetric. 
2) Covariant energy conservation $\nabla_{\mu}T^{\mu\nu}=0$. 
3) Possesses Schwarzschild\textendash{}like features $g_{00}=-g_{rr}^{-1}$. 
As a result, 
the matter source distribution's spherically symmetric energy\textendash{}momentum tensor \cite{bib17,bib56} can be expressed as $T_{\mu\nu}=(\rho+p_{\vartheta_{i}})U_{\mu}U_{\nu}+p_{\vartheta_{i}}g_{\mu\nu}+(p_{r}-p_{\vartheta_{i}})X_{\mu}X_{\nu}$. 
where $U_{\mu}$ is the fluid velocity and $X_{\mu}$ is the unit vector along the radial direction,
\begin{equation} \label{eq4}
[T^{\mu}_{\quad\nu}]_{\text{matter}}=diag[-\rho_{\text{matter}}(r),p_{r},p_{\vartheta_{1}},\dots,p_{\vartheta_{D-2}}],
\end{equation}
where $p_{r}=-\rho_{\text{matter}}(r)$ denotes radial pressure, 
while $p_{\vartheta_{i}}=-\rho_{\text{matter}}(r)-[r/(D-2)]\partial_{r}\rho_{\text{matter}}(r)$ denotes tangential pressure.
The Einstein field equation can be written as
\begin{equation} \label{eq5}
R_{\mu\nu}-\frac{1}{2}Rg_{\mu\nu}=8\pi G_{D}[T_{\mu\nu}]_{\text{matter}},
\end{equation}
where $G_{D}$ is the gravitational constant in $D$\textendash{}dimensional spacetime, 
$G_{D}=m^{-2}_{\text{Pl}}=l^{-2}_{\text{Pl}}$, 
where $m^{-2}_{\text{Pl}}$ and $l^{-2}_{\text{Pl}}$ represent the Planck mass and length, respectively.

In the $D$\textendash{}dimensional Schwarzschild\textendash{}Tangherlini spacetime, 
the metric of a non\textendash{}commutative spherically symmetric black hole is
\begin{equation} \label{eq6}
\mathrm{d}s^{2}=-f(r)\mathrm{d}t^{2}+f(r)^{-1}\mathrm{d}r^{2}+r^{2}\mathrm{d}\Omega^{2}_{D-2},
\end{equation}
where $\mathrm{d}\Omega^{2}_{D-2}$ denotes the line element on the $(D-1)$ dimensional unit sphere, 
and
\begin{eqnarray} \label{eq7}
\mathrm{d}\Omega^{2}_{D-2}=&&\mathrm{d}\vartheta_{1}^{2}+\sin^{2}\vartheta_{1}\mathrm{d}\vartheta_{2}^{2}+\dots\nonumber\\
&&+(\sin^{2}\vartheta_{1}\dots\sin^{2}\vartheta_{D-3})\mathrm{d}\vartheta_{D-2}^{2}\nonumber\\
=&&\sum_{j=1}^{D-2}\left(\prod_{i=1}^{j=2}\sin^{2}\vartheta_{i}\right)\mathrm{d}\vartheta_{j}^{2},
\end{eqnarray}
here, 
the coordinate system $(t,r,\vartheta_{1},\vartheta_{2},\dots,\vartheta_{D-2})$ is utilized. 
The lapse function \cite{bib21} is
\begin{equation} \label{eq8}
f(r)=1-\frac{16\pi G_{d}m(r)}{(D-2)\Omega_{D-2}r^{D-3}},
\end{equation}
where $\Omega_{D-2}$ denotes the volume of a $(D-2)$\textendash{}dimensional unit sphere \cite{bib57,bib58}.
\begin{equation} \label{eq9}
\Omega_{D-2}=\frac{2\pi^{\frac{D-1}{2}}}{\Gamma\left(\frac{D-1}{2}\right)}.
\end{equation}
$m(r)=\int_{0}^{+\infty}\rho_{\text{matter}}(r)\Omega_{D-2}r^{2}\mathrm{d}r$ denotes the black hole's mass distribution, 
which is given by
\begin{equation} \label{eq10}
m(r)=\frac{M}{\Gamma\left(\frac{D+k-1}{2}\right)}\gamma\left(\frac{D+k-1}{2},\frac{r^{2}}{4\theta}\right).
\end{equation}

This enables us to rewrite the lapse function as follows:
\begin{equation} \label{eq11}
f(r)=1-\frac{16\pi\frac{M}{\Gamma\left(\frac{D+k-1}{2}\right)}}{(D-2)\Omega_{D-2}r^{D-3}}\cdot\gamma\left(\frac{D+k-1}{2},\frac{r^{2}}{4\theta}\right).
\end{equation}
$\gamma(a,x)$ is the lower incomplete gamma function. 
Geometric units are utilized throughout the rest of this text such that $G_{D}=c=\hbar=k_{B}=1$.

The next introduce numerous special black hole solutions, which are variants of non\textendash{}commutative $D$\textendash{}dimensional Schwarzschild black hole solutions with generic smeared matter distribution. 
Eq. (\ref{eq11}) is close to the lapse function of a $D$\textendash{}dimensional Schwarzschild black hole \cite{bib59} when the non\textendash{}commutative limit $\theta\to0$ or $r\gg\theta$ is used. 
Eq. (\ref{eq11}) depicts the non\textendash{}commutative $D$\textendash{}dimensional Schwarzschild black hole \cite{bib15} under Gaussian smeared matter distribution when $k=0$ is met. 
When both of the above requirements are satisfied, 
Eq. (\ref{eq11}) turns into a Schwarzschild black hole.

\subsection{Perturbation equation and effective potential for scalar field} \label{subsec2.2}

For the  mathematical black holes,
we only discuss the scattering of massless scalar waves by black holes.
This is the most basic scenario so that we can easily grasp some properties about black hole.
Although this study has mostly focused on massless scalar wave fields, 
identical strategies may be used to any scattering issue.
The Klein\textendash{}Gordon equation in the spacetime background Eq. (\ref{eq6}) describes the massless scalar wave in this situation.
\begin{equation} \label{eq12}
(-g)^{-\frac{1}{2}}\partial_{\mu}[(-g)^{\frac{1}{2}}g^{\mu\nu}\partial_{\nu}\Psi]=0,
\end{equation}
where $g$ denotes the determinant of $g_{\mu\nu}$, 
$\Psi$ represents a scalar field \cite{bib60}.
\begin{equation} \label{eq13}
\Psi(t,r,{\vartheta_{i}})=\sum_{l,m}r^{\frac{2-D}{2}}\psi_{l}(t,r)Y_{l,m}(\vartheta_{1},\vartheta_{2},\dots,\vartheta_{D-2}),
\end{equation}
where ${\vartheta_{i}}$ denotes the $(D-2)$\textendash{}dimensional angle and $Y_{l,m}(\vartheta_{1},\vartheta_{2},\dots,\vartheta_{D-2})$ denotes the $D$\textendash{}dimensional spherical harmonic function $\psi_{l}(t,r)=e^{i\omega t}\Phi_{\omega}(r)$. 
The Klein\textendash{}Gordon equation may be reduced to a Schr\"{o}dinger\textendash{}like wave function using the tortoise coordinate transform $r_{*}=\int f(r)^{-1}\mathrm{d}r$ and the separation of radial and angular variables. 
\begin{equation} \label{eq14}
\left[\frac{\partial^{2}}{\partial r_{*}^{2}}+\omega^{2}-V_{\theta,k}(r_{*})\right]\Phi_{\omega}=0.
\end{equation}
The effective potential is
\begin{eqnarray} \label{eq15}
V_{\theta,k}(r)=&&f(r)\left[\frac{l(l+D-3)}{r^{2}}+\frac{D-2}{2r}\frac{\partial f(r)}{\partial r}\right]\nonumber\\
&&+f(r)\left[\frac{(D-2)(D-4)}{4r^2}f(r)\right],
\end{eqnarray}
where $l=1,2,\dots$ is a multipole number. 
The tortoise coordinate $r_{*}$ is defined on the interval $(-\infty,+\infty)$ such that $r_{*}\to+\infty$ relates to the spatial infinite $r\to+\infty$ and $r_{*}\to-\infty$ corresponds to the event horizon $r_{\text{eh}}$. 
The above effective potential is defined positively and takes the form of a potential barrier close to zero at both spatial infinity and the event horizon. 
We discover that as the parameters $k$ or $D$ grows, 
so does the maximum value of the effective potential barrier $V_{\text{max}}$. 
And when the parameter $\theta$ grows, 
the maximum value of the potential barrier $V^{\text{max}}_{\theta,k}(r)$ lowers. 
The influence of the parameter $\theta$ falls to practically null as the value of $k$ grows.


\subsection{The allowable value of $k$ and the valid range of $\theta$} \label{subsec2.3}

According to Eq. (\ref{eq11}), 
the formula for $M$ and the event horizon radius $r_{\text{eh}}$ can be obtained.
\begin{equation} \label{eq16}
M=\frac{(D-2)\Omega_{D-2}r_{\text{eh}}^{D-3}}{16\pi}\cdot\frac{\Gamma\left(\frac{D+k-1}{2}\right)}{\gamma\left(\frac{D+k-1}{2},\frac{r_{\text{eh}}^{2}}{4\theta}\right)},
\end{equation}
where $r_{\text{eh}}$ is the greatest root of $f(r)=0$. 
The radius of an extreme black hole's event horizon $r_{\text{eh}}$ satisfies the relation $\partial M/\partial r_{\text{eh}}=0$, 
therefore, 
$r_{\text{eh}}$ satisfies the following formula,
\begin{equation} \label{eq17}
2x_{\text{eh}}^{D+k-1}e^{-x_{\text{eh}}^{2}}=(D-3)\cdot\gamma\left(\frac{D+k-1}{2},x_{\text{eh}}^{2}\right),
\end{equation}
where $x_{\text{eh}}$ is defined by $x_{\text{eh}}=r_{\text{eh}}/\sqrt{4\theta}$.

The hoop conjecture \cite{bib61,bib62} needs to be considered here: 
The mean radius of the matter mass distribution is smaller than the radius of the extreme black hole's event horizon. 
This is $\bar{r}\le r_{\text{eh}}$ or $\bar{x}\le x_{\text{eh}}$, 
where $\bar{x}_{\text{eh}}=\bar{r}_{\text{eh}}/\sqrt{4\theta}$. 
Furthermore, 
the hoop conjecture ensures that the smallest black holes (extreme black holes) have zero temperature and zero heat capacity. 
Black holes will not have extreme structures if the hoop conjecture is broken. 
The following equation is obtained by combining Eq. (\ref{eq3}).
\begin{equation} \label{eq18}
\frac{\Gamma\left(\frac{D+k}{2}\right)}{\Gamma\left(\frac{D+k-1}{2}\right)}\le x_{\text{eh}}(D,k).
\end{equation}
As a result, 
the permissible range of $k$ of the non-commutative $D$\textendash{}dimensional Schwarzschild\textendash{}Tangherlini black hole corresponding to different $D$ values can be determined using Eq. (\ref{eq18}), 
as shown in TABLE \ref{tab1}. 
The Gaussian mass distribution of matter is only valid to non\textendash{}commutative Schwarzschild\textendash{}Tangherlini black holes in $D=4$ and $D=5$ spacetime, 
as shown in TABLE \ref{tab1}.

\begin{table}[t]
\caption{\label{tab1}
The allowable range of $k$ values corresponding to different $D$ values.
}
\begin{ruledtabular}
\begin{tabular}{ccccccccc}
$D$ & 4         & 5         & 6         & 7          & 8           & 9           & 10         & 11         \\ \hline
$k$ & $\ge0$ & $\ge0$ & $\ge4$ & $\ge8$ & $\ge14$ & $\ge22$ & $\ge32$ & $\ge43$ \\
\end{tabular}
\end{ruledtabular}
\end{table}

\begin{figure}[b]
\includegraphics[width=0.45\textwidth]{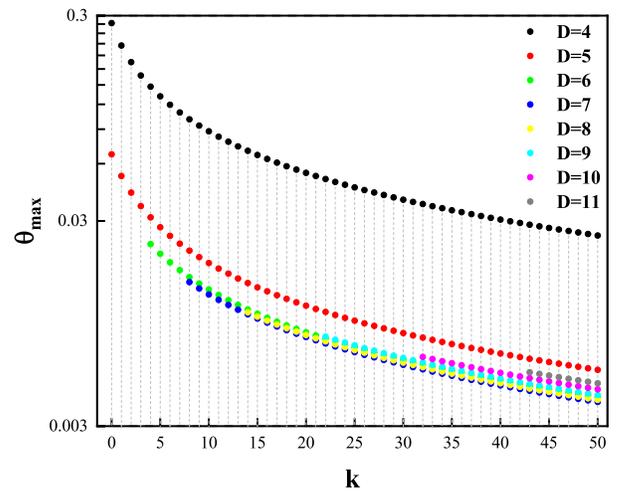}
\caption{\label{fig1}$\theta_{\text{max}}$ values for different $k$ and $D$ values.
}
\end{figure}

\begin{table}[t]
\caption{\label{tab2}
$\theta_{\text{max}}$ values for different $k$ and $D$ values.
}
\begin{ruledtabular}
\begin{tabular}{ccccccc}
$D$ & $k=0$          & $k=1$      & $k=2$      & $k=3$      & $k=4$      \\ 
4     & 0.275811     & 0.214662   & 0.17807    & 0.153293   & 0.135218   \\ 
5     & 0.0633279   & 0.0496268  & 0.0412099  & 0.035442   & 0.0312102  \\ \hline
$D$ & $k=4$         & $k=5$      & $k=6$      & $k=7$      & $k=8$       \\  
6     & 0.0231343   & 0.0207527  & 0.0188472  & 0.0172844  & 0.0159774  \\ \hline
$D$ & $k=8$         & $k=9$      & $k=10$     & $k=11$     & $k=12$    \\ 
7     & 0.0150971   & 0.0140589  & 0.0131622  & 0.0123793  & 0.0116894  \\ \hline
$D$ & $k=14$       & $k=15$     & $k=16$     & $k=17$     & $k=18$    \\ 
8     & 0.0107588   & 0.0102581  & 0.00980412 & 0.00939053 & 0.00901203 \\ \hline
$D$ & $k=22$       & $k=23$     & $k=24$     & $k=25$     & $k=26$     \\ 
9     & 0.00818131  & 0.007913   & 0.00766244 & 0.00742791 & 0.00720788 \\ \hline
$D$ & $k=32$        & $k=33$     & $k=34$     & $k=35$     & $k=36$    \\ 
10   & 0.00652586  & 0.00636871 & 0.00621922 & 0.00607684 & 0.00594106 \\ \hline
$D$ & $k=43$       & $k=44$     & $k=45$     & $k=46$     & $k=47$     \\ 
11   & 0.0054962   & 0.0053935  & 0.00529468 & 0.00519954 & 0.00510786 \\ 
\end{tabular}
\end{ruledtabular}
\end{table}

It is worth mentioning that the lapse function $f(r)=0$ in a $D$\textendash{}dimensional Schwarzschild\textendash{}Tangherlini black hole can have 0, 1, or 2 horizons depending on the value of $\theta$. 
As a result, 
we may obtain the extreme $\theta$ parameter of the coincidence of the inner and outer horizon, 
denoted as $\theta_{\text{max}}$. 
We compute $\theta_{\text{max}}$ for the permitted $k$ of $4\le D\le11$ spacetime, 
as indicated in  FIG. \ref{fig1}, 
and present the $\theta_{\text{max}}$ for the six least allowable $k$ in TABLE \ref{tab2} for clarity. 
We discover that when $4\le D\le11$, 
$\theta_{\text{max}}$ decreases as the value of $k$ grows.

\subsection{Shadow of $4-D$ cases} \label{subsec2.4}

We discuss the shadow in $4-D$ non\textendash{}commutative Schwarzschild\textendash{}Tangherlini black hole spacetimes in this subsection. 
In general, 
the geodesics in black hole spacetimes are described using the formula \cite{bib47}
\begin{equation} \label{eq19}
\dot{s}^{2}=-f(r)\dot{t}^{2}+f^{-1}(r)\dot{r}^{2}+r^{2}(\dot{\theta}^{2}+\sin^{2}\theta\dot{\phi}^{2})=\lambda,
\end{equation}
where the over dot denotes the derivative with respect to an affine parameter $\lambda$. For massless particles we have $\lambda = 0 $.

It is useful to introduce the function $h(r)^2$
\begin{equation} \label{eq20}
h(r)^{2}=\frac{r^{2}}{f(r)},
\end{equation}
In terms of $h(r)^2$, the radius of the outermost photon sphere $r_{\rm{ph}}$ is the largest root of the equation \cite{bib63}
\begin{equation} \label{eq21}
\frac{\mathrm{d}}{\mathrm{d}r}h(r)^{2}=0,
\end{equation}

The critical impact parameter $b_{\rm{cr}}$ and the radius of the photon sphere $r_{\rm{ph}}$ are related by
\begin{equation} \label{eq22}
b_{\rm{cr}}=h(r_{\rm{ph}}).
\end{equation}

The shadow radius $r_{\rm{sh}}$ is given by
\begin{equation} \label{eq23}
r_{\rm{sh}}=\left.\frac{r}{\sqrt{f{(r)}}}\right\vert_{r_{\rm{ph}}}.
\end{equation}

The constraints utilized in Ref. \cite{bib68} will be used to limit the parameters of $4-D$ non\textendash{}commutative Schwarzschild\textendash{}Tangherlini black hole in the next parts.
The $1\sigma$ constraints

\begin{equation} \label{eq24}
4.54\lesssim r_{\rm{sh}}/M\lesssim5.22,
\end{equation}
and
\begin{equation} \label{eq25}
4.20\lesssim r_{\rm{sh}}/M\lesssim5.56.
\end{equation}
These constraints are consistent with those mentioned in Ref. \cite{bib69}
These estimates come from two completely independent observation,
Keck-based and VLTI-based.

\begin{figure}[t]
\includegraphics[width=0.45\textwidth]{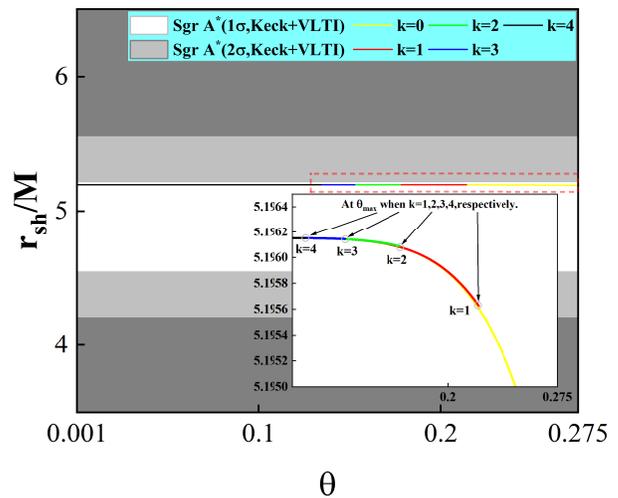}
\caption{\label{fig2}Shadow radius $r_{sh}$ of $4-D$ non\textendash{}commutative Schwarzschild\textendash{}Tangherlini black hole as a function of the non\textendash{}commutative parameter $\theta$($k=0,1,2,3,4$ corresponding to the Yellow, red, green, blue and black solid lines, respectively).
The value of $\theta$ starts from $\theta\to0$.
The white and light gray zones match the EHT horizon-scale picture of Sgr A* at 1$\sigma$ and 2$\sigma$, respectively.
The gray area represents the value outside the limit.
}
\end{figure}

We plot the shadows for $4-D$ non\textendash{}commutative Schwarzschild\textendash{}Tangherlini black hole as a function of the non\textendash{}commutative parameter $\theta$ in Fig. \ref{fig2} and for $k=0,1,2,3,4$.
Then we find that it makes very little effect on black hole shadow whether it is $k$ or $\theta$.
We only consider the $4-$dimensional case here since the estimates for these observations are based on the $4-$dimensional GR, 
and we can't easily extend them to the high-dimensional situation.

Because of the existence of the innermost unstable photon sphere $r=r_{\rm{ph}}$, 
the photon can deviate from the photon sphere at any angle. 
This feature indicates that the glory phenomena will occur \cite{bib47}. 
Glory phenomena, 
like those seen in optics, 
are bright spots or halos that occur in the scattering intensity in the epipolar direction, 
the intensity and size of which depend on the wavelength of the incident perturbation. 
The approximation developed by Matzner et al. \cite{bib62} may be used to determine the magnitude and size of bright spots.

\begin{figure*}
\includegraphics[width=0.3\textwidth]{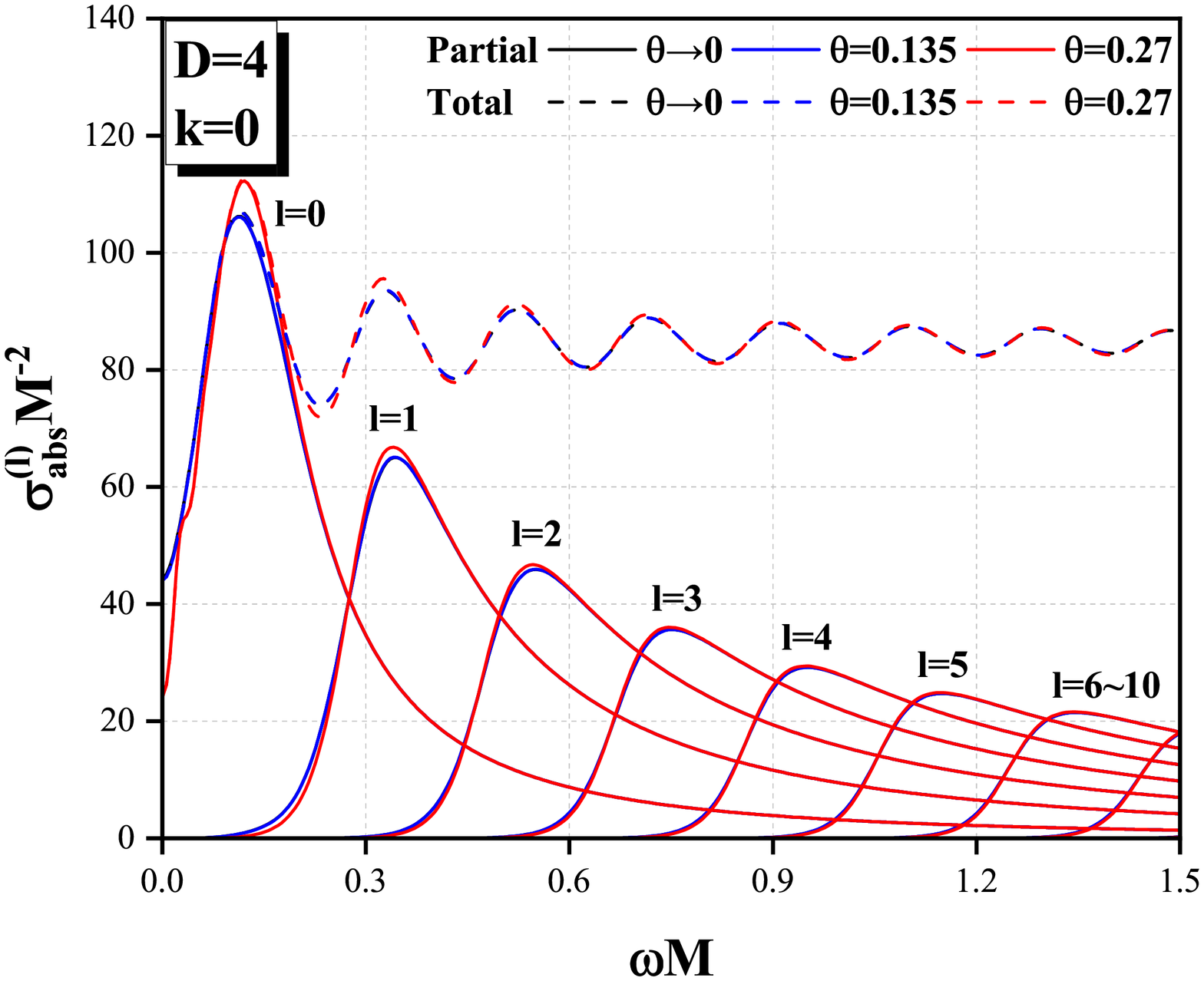}
\includegraphics[width=0.3\textwidth]{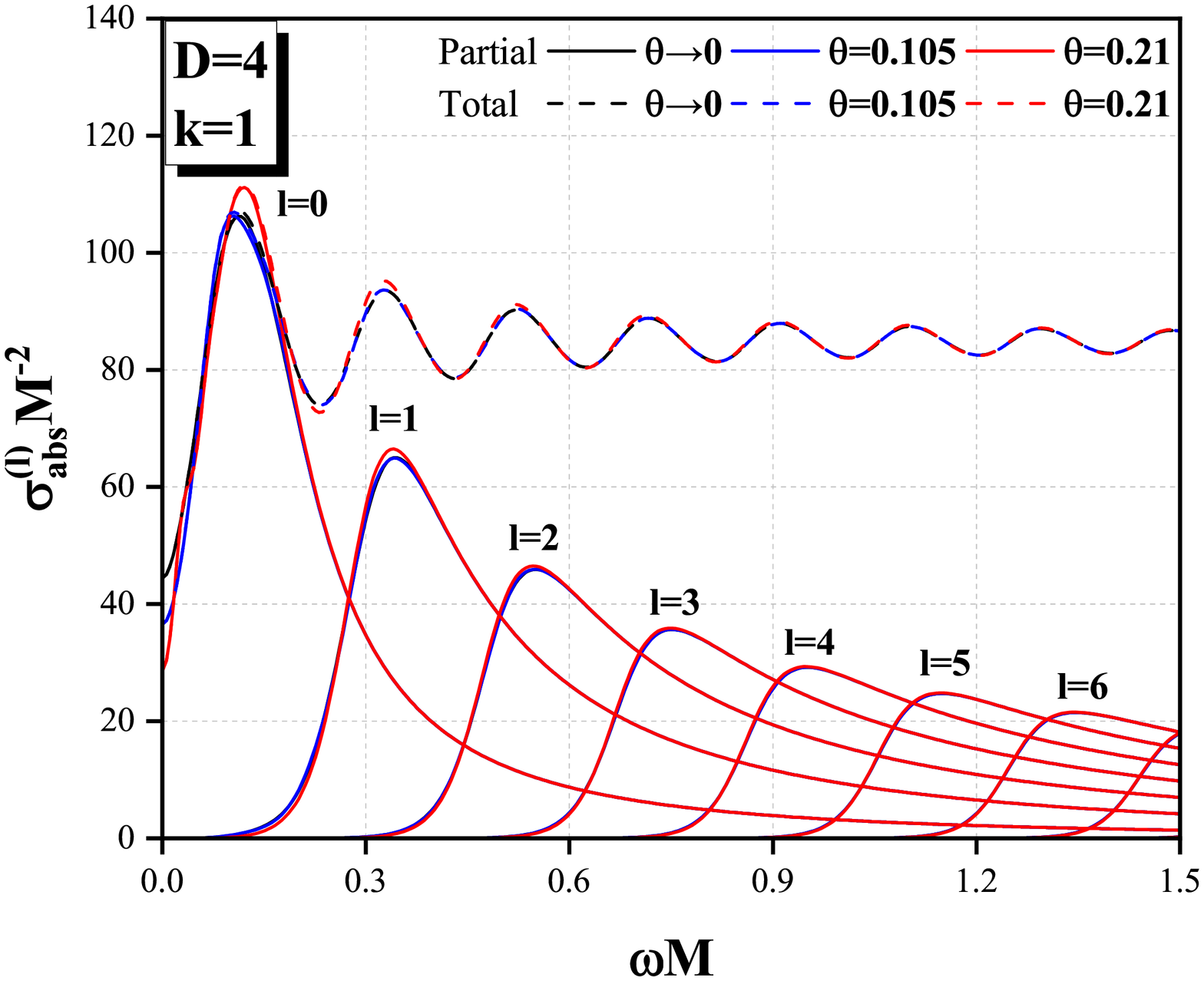}
\includegraphics[width=0.3\textwidth]{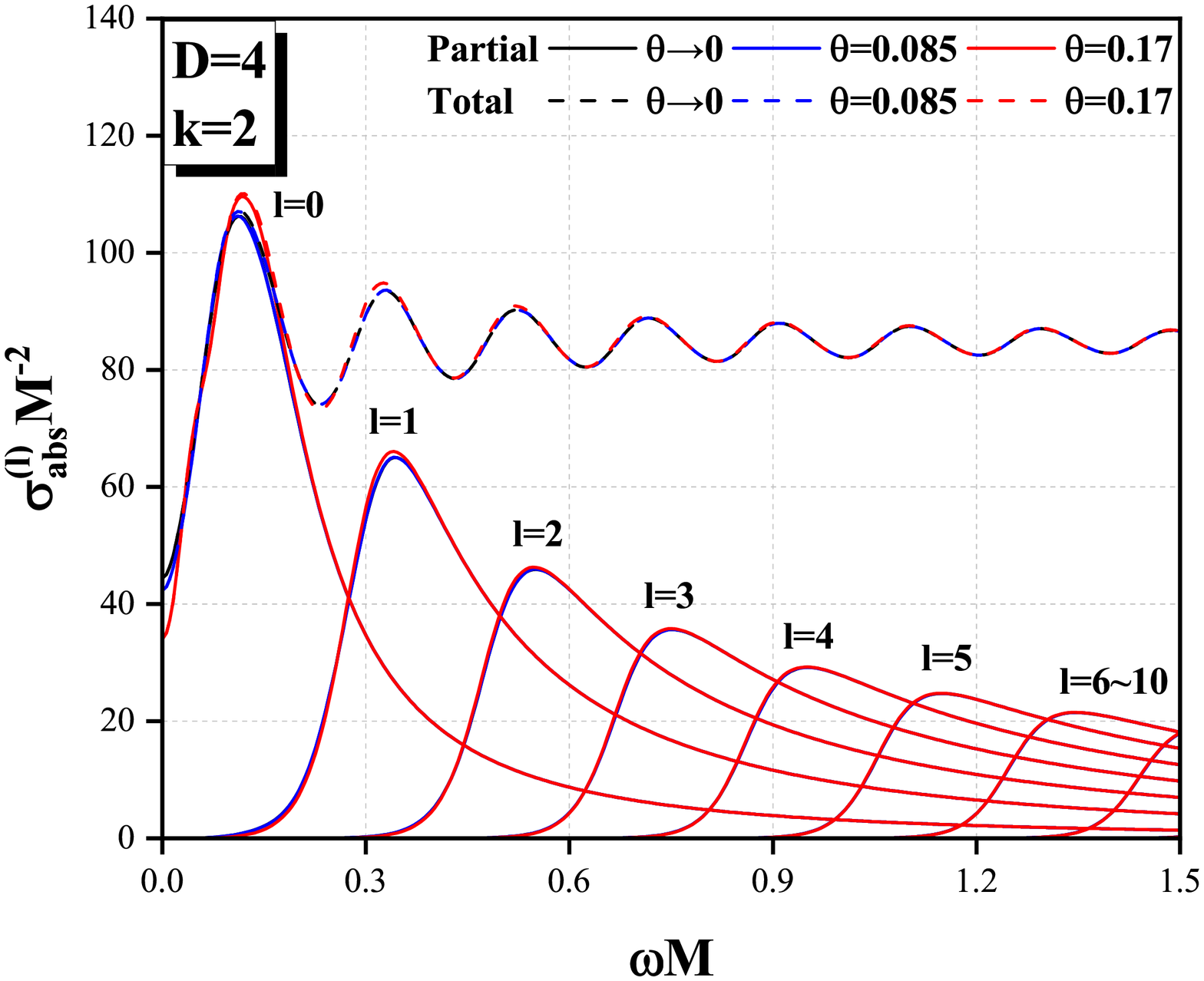}
\includegraphics[width=0.3\textwidth]{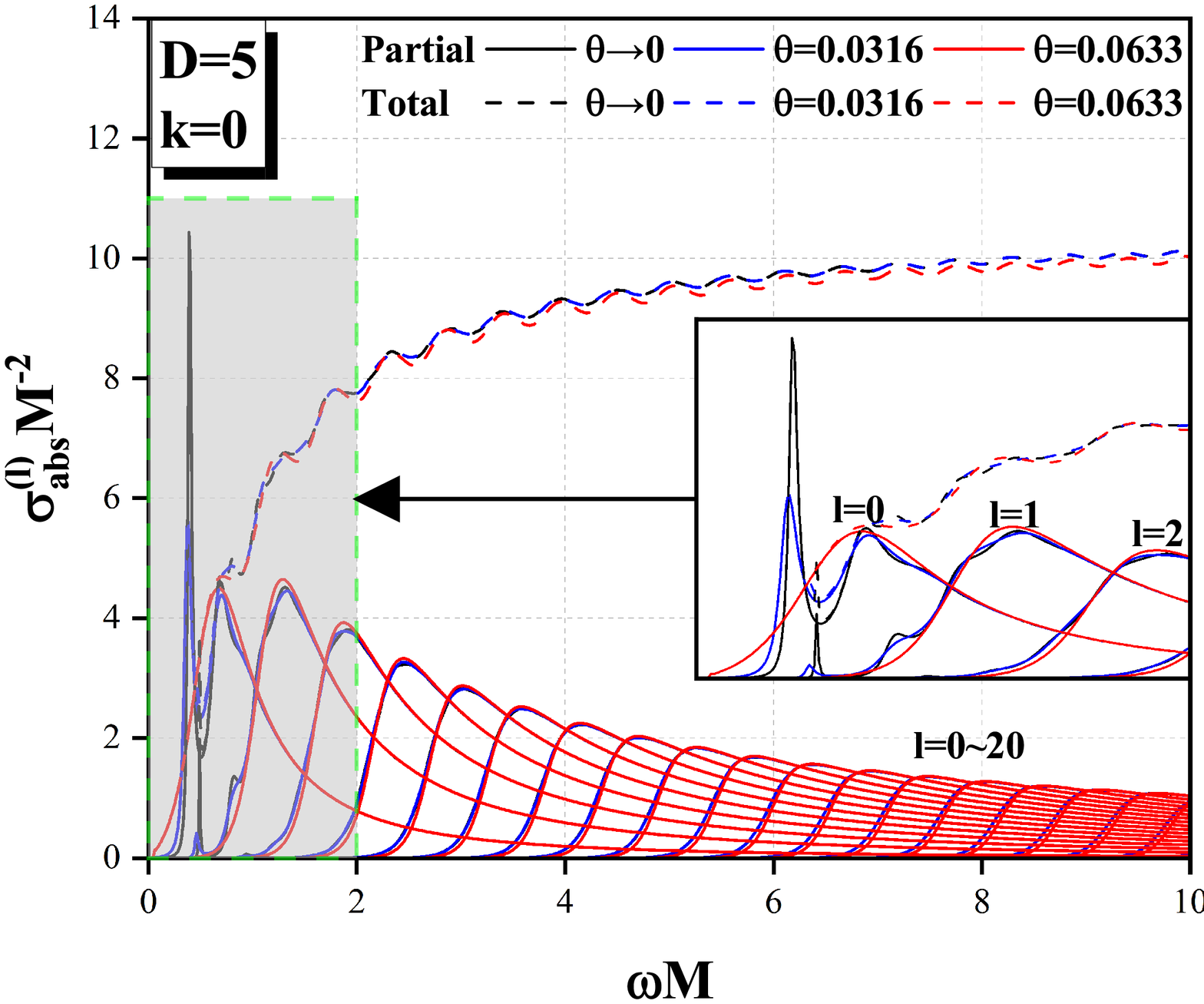}
\includegraphics[width=0.3\textwidth]{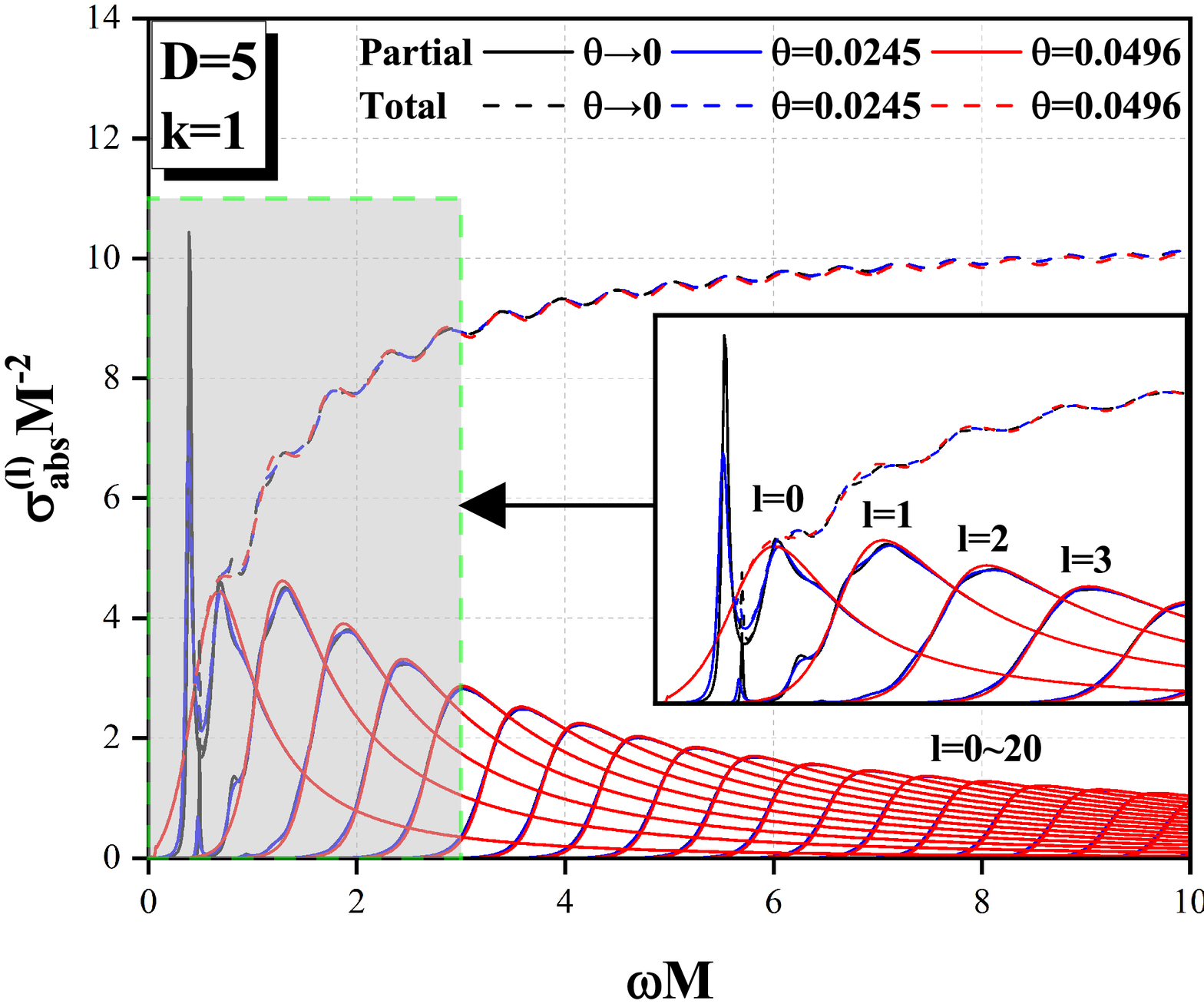}
\includegraphics[width=0.3\textwidth]{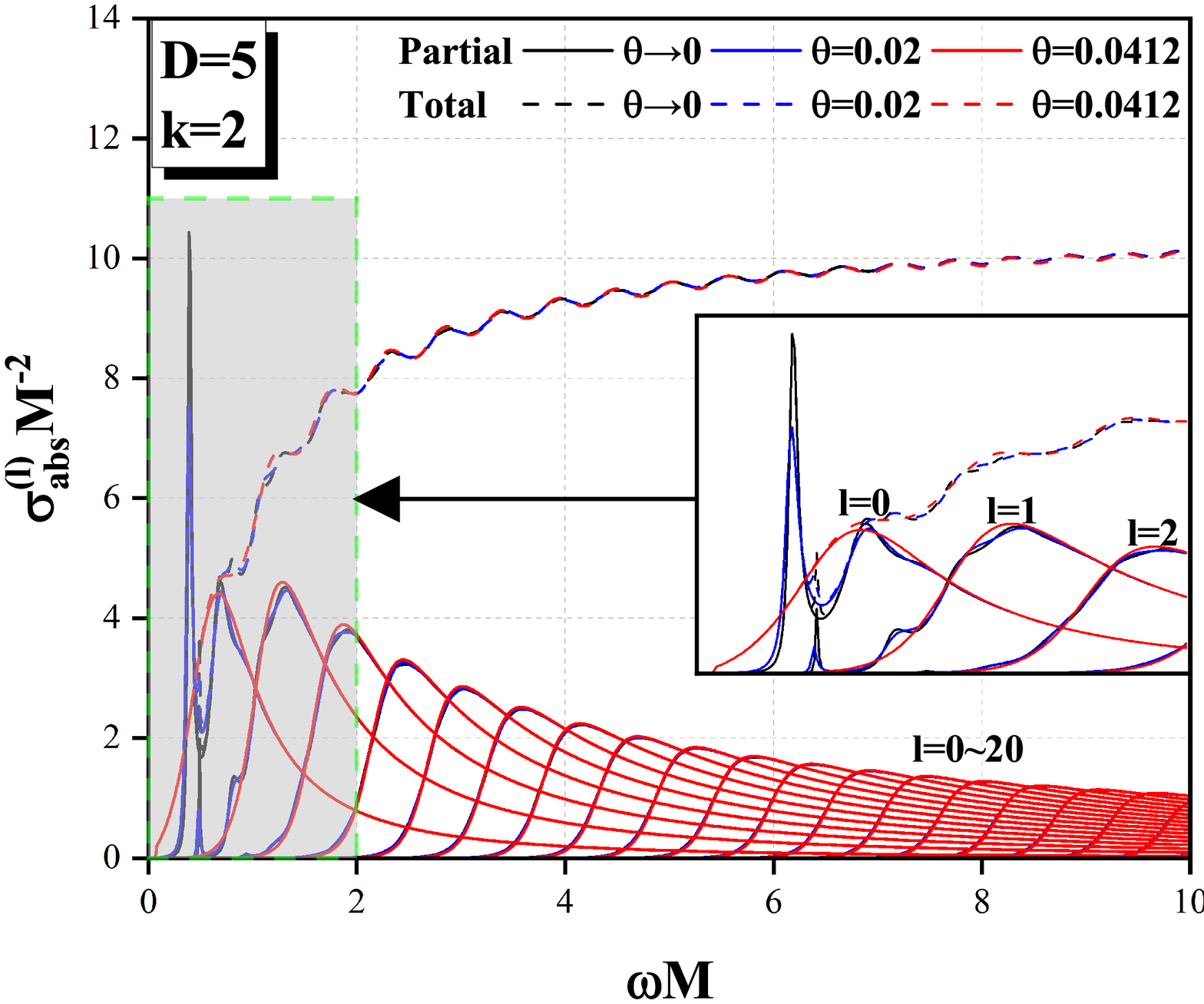}
\includegraphics[width=0.3\textwidth]{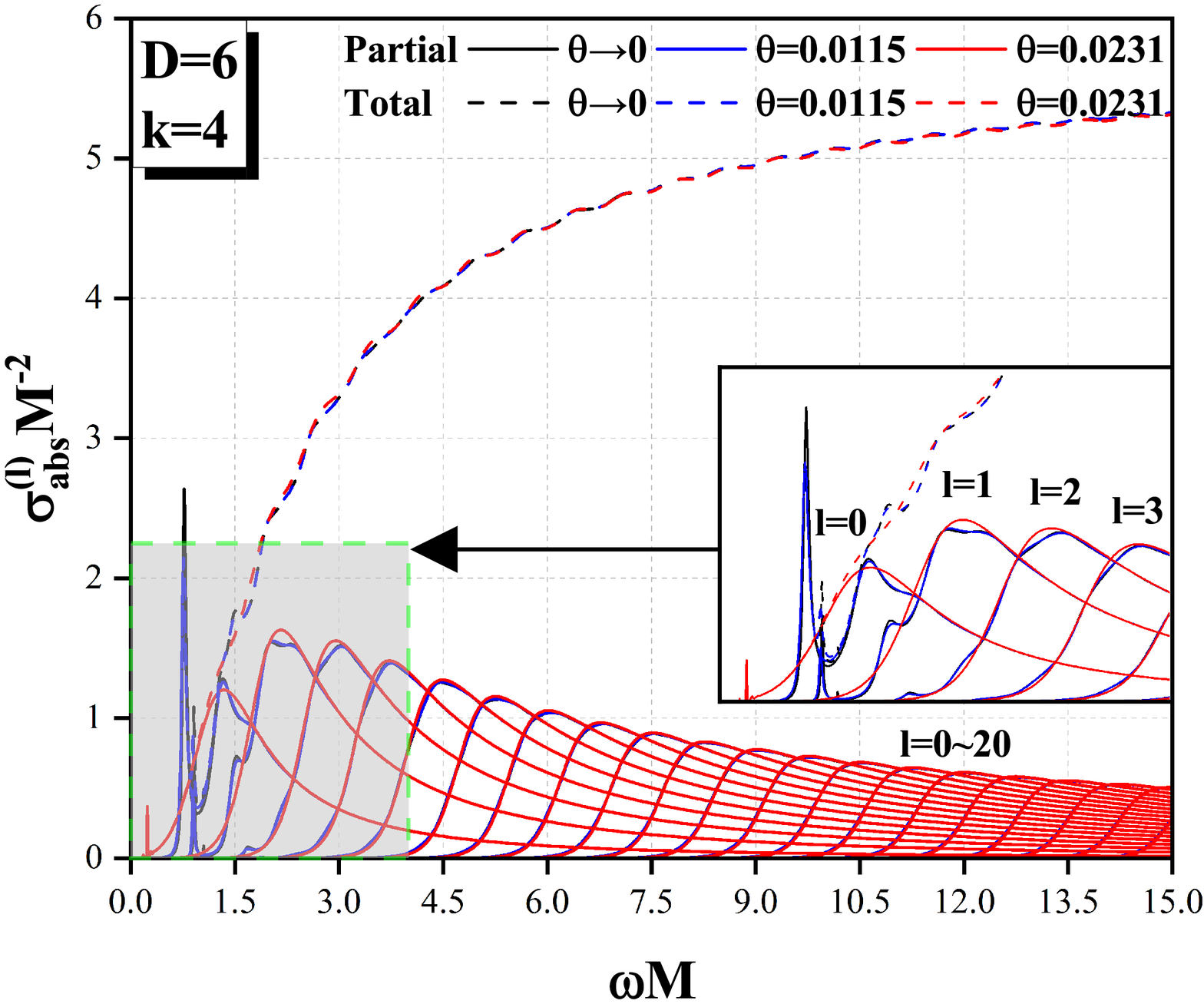}
\includegraphics[width=0.3\textwidth]{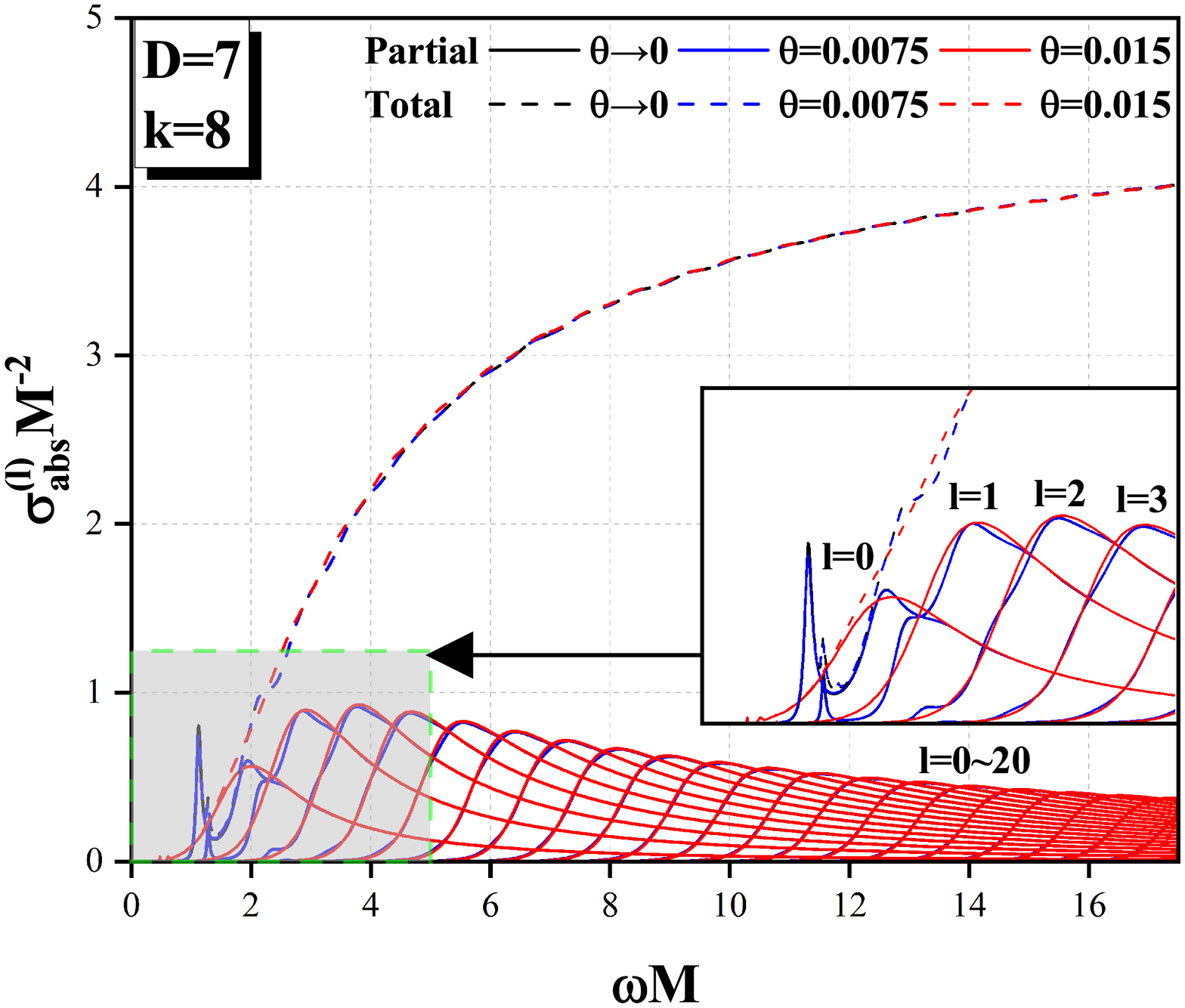}
\includegraphics[width=0.3\textwidth]{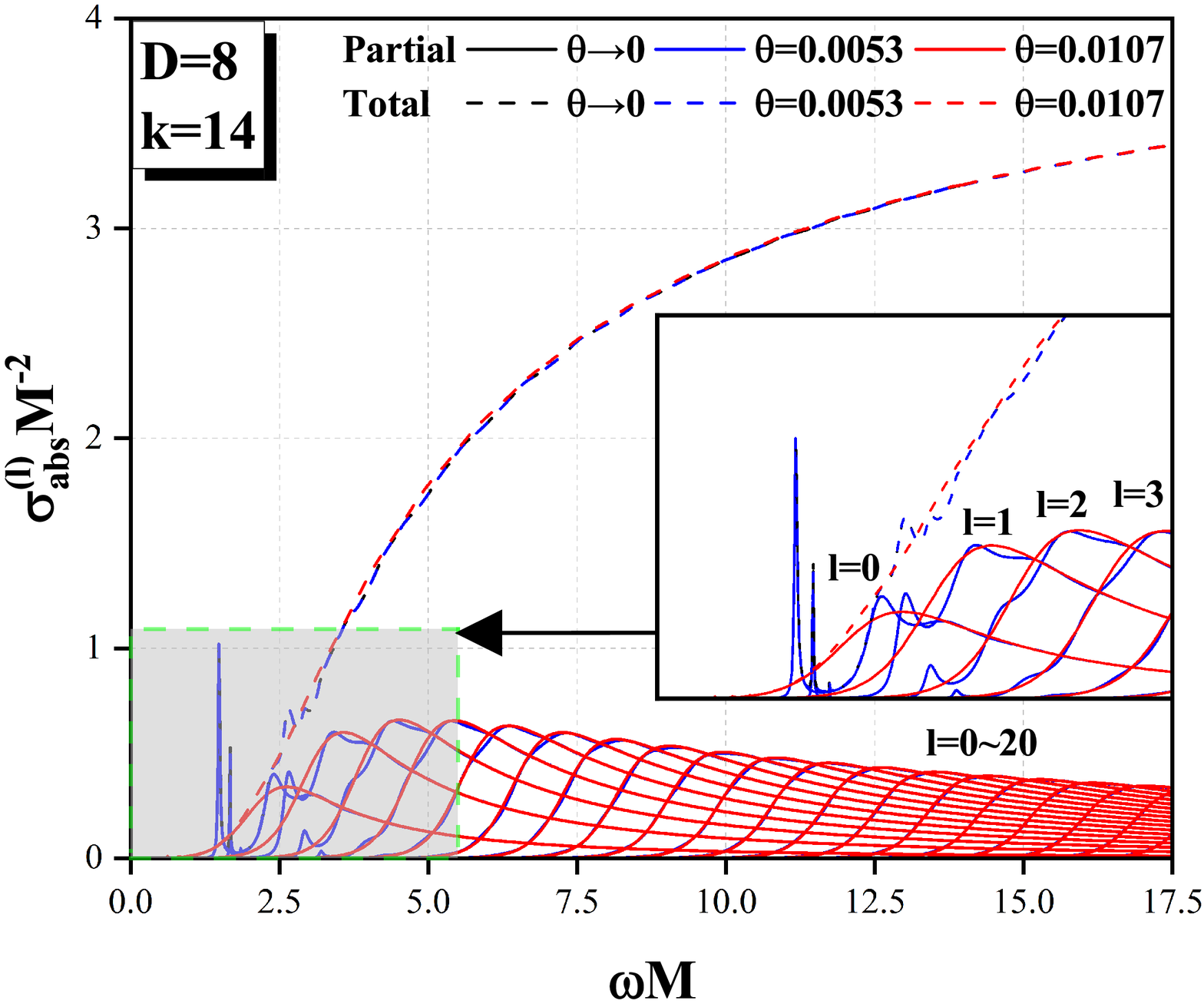}
\includegraphics[width=0.3\textwidth]{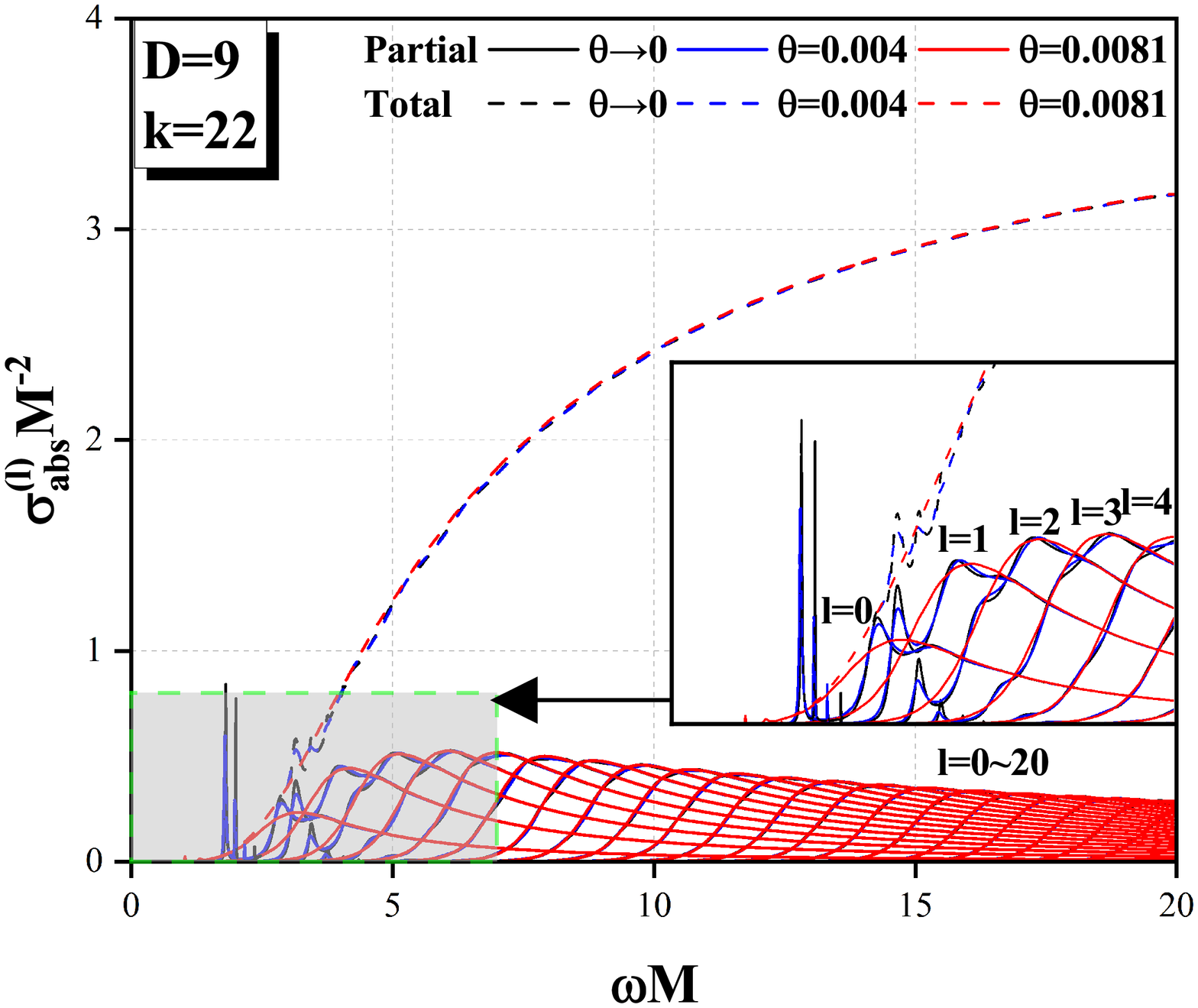}
\includegraphics[width=0.3\textwidth]{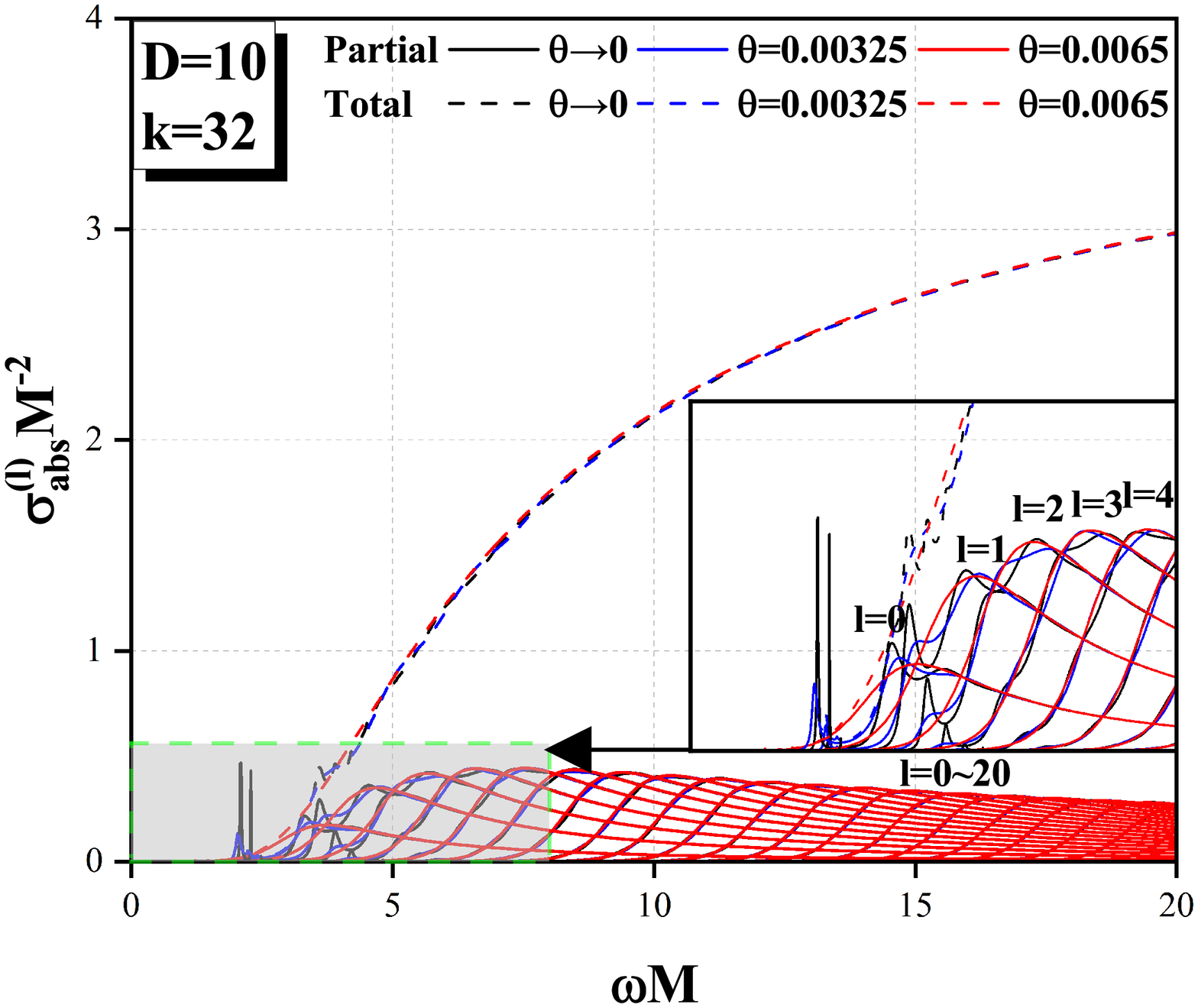}
\includegraphics[width=0.3\textwidth]{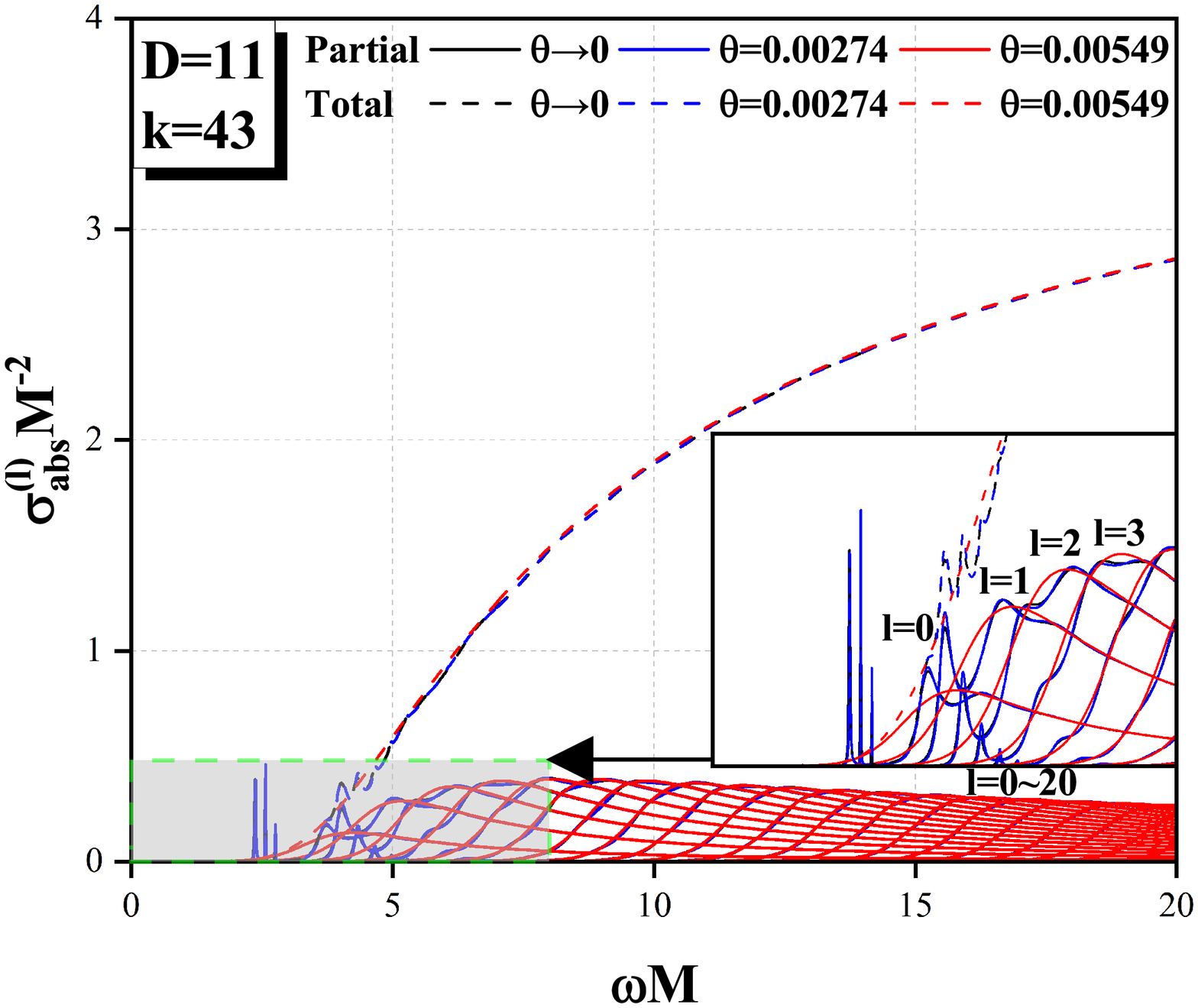}
\caption{\label{fig3}The partial absorption cross section and total absorption cross section of high dimensional non\textendash{}commutative black holes with different values of $k$ and $D$ are plotted respectively, 
and the mass is normalized to M=1. 
The parameter $\theta$ has a slight effect on the absorption cross section.
}
\end{figure*}

\begin{figure*}
\includegraphics[width=0.3\textwidth]{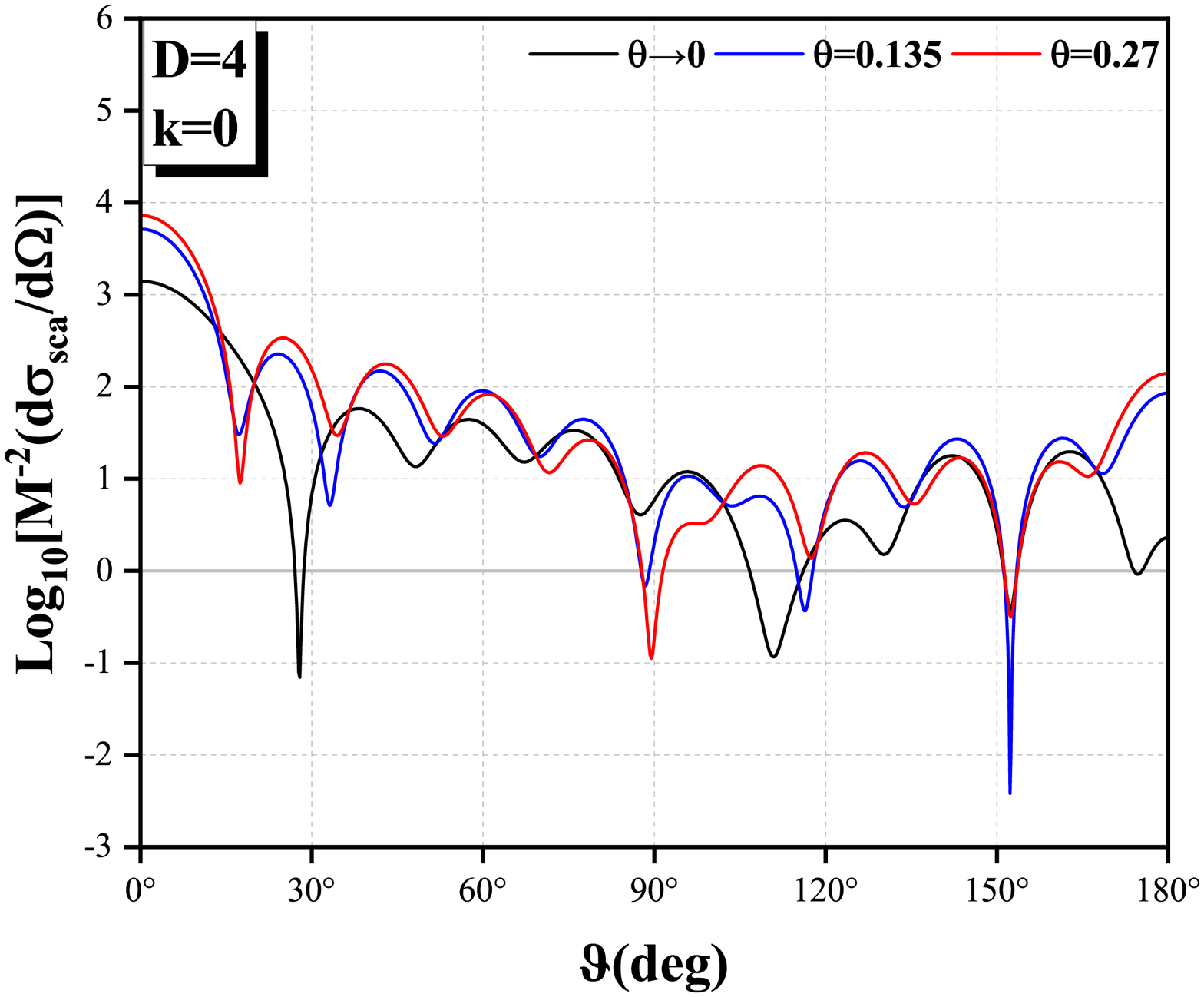}
\includegraphics[width=0.3\textwidth]{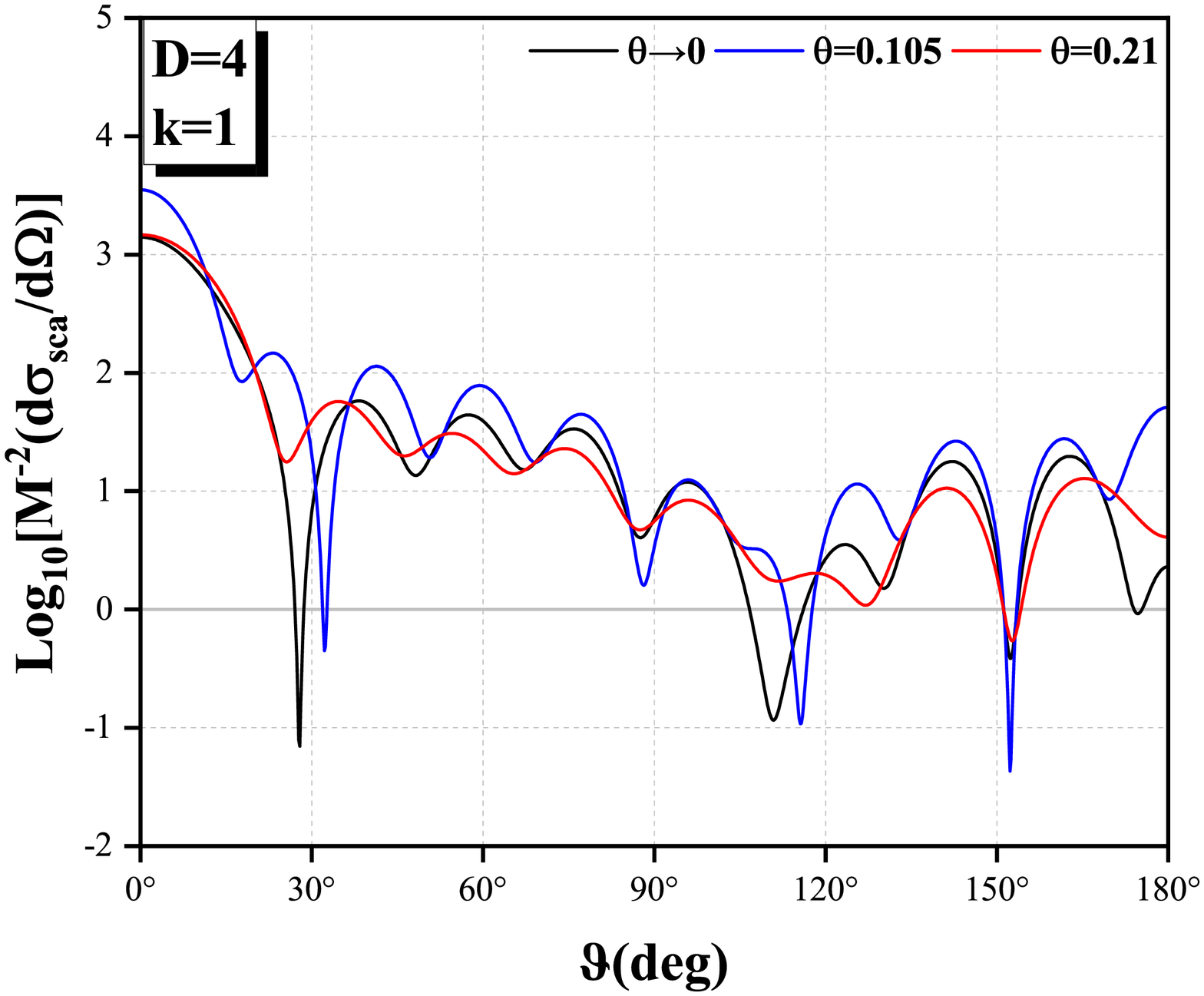}
\includegraphics[width=0.3\textwidth]{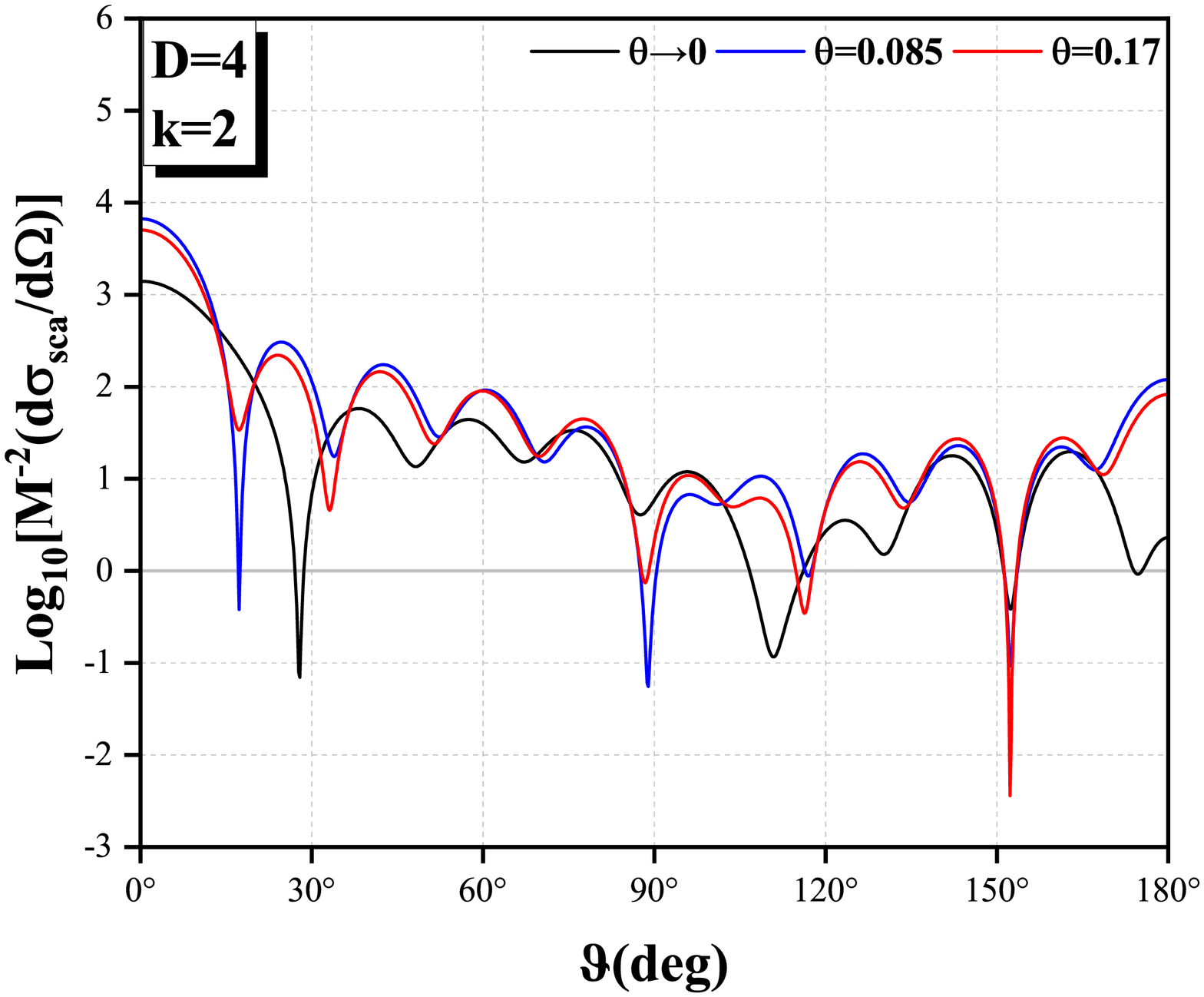}
\includegraphics[width=0.3\textwidth]{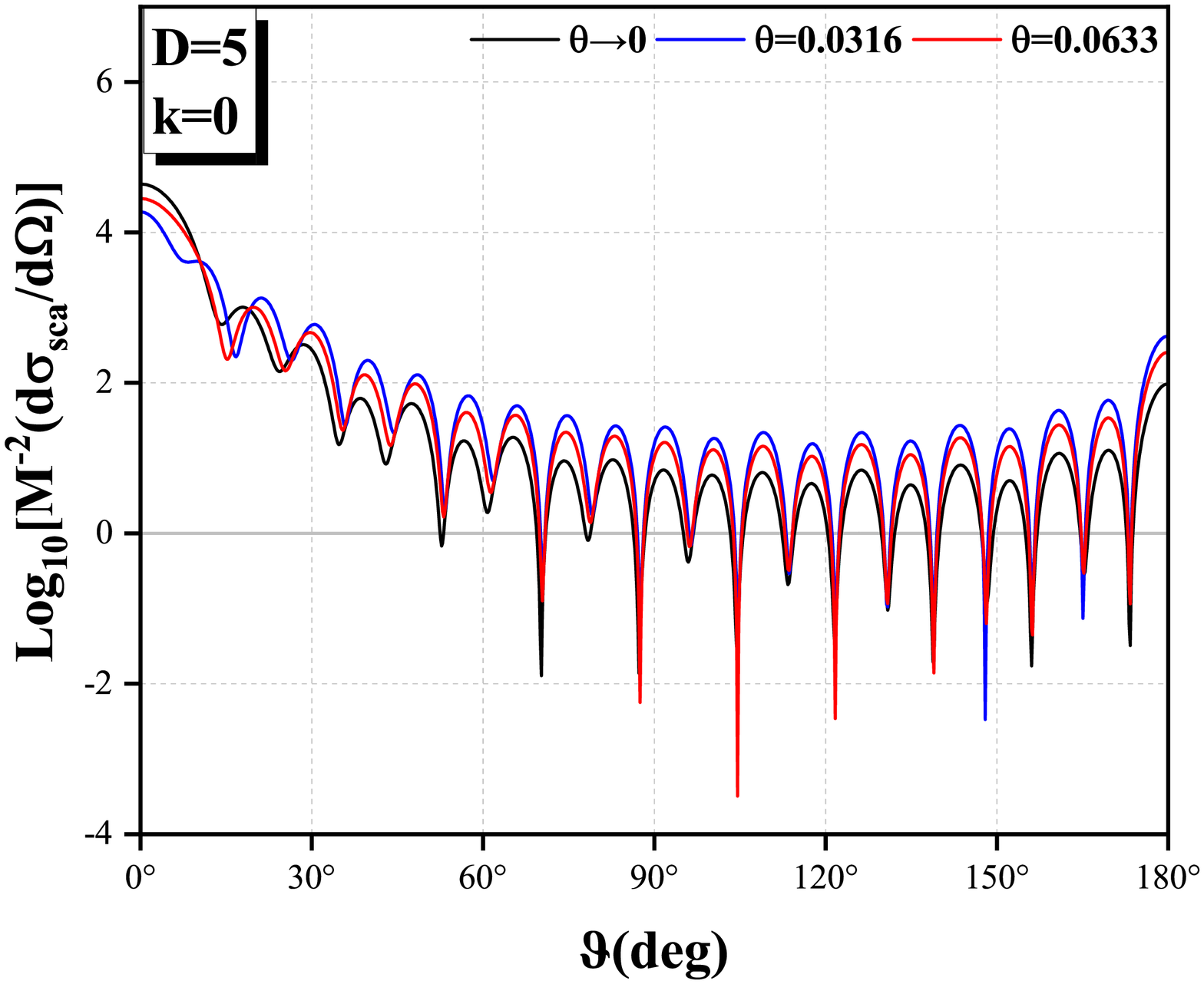}
\includegraphics[width=0.3\textwidth]{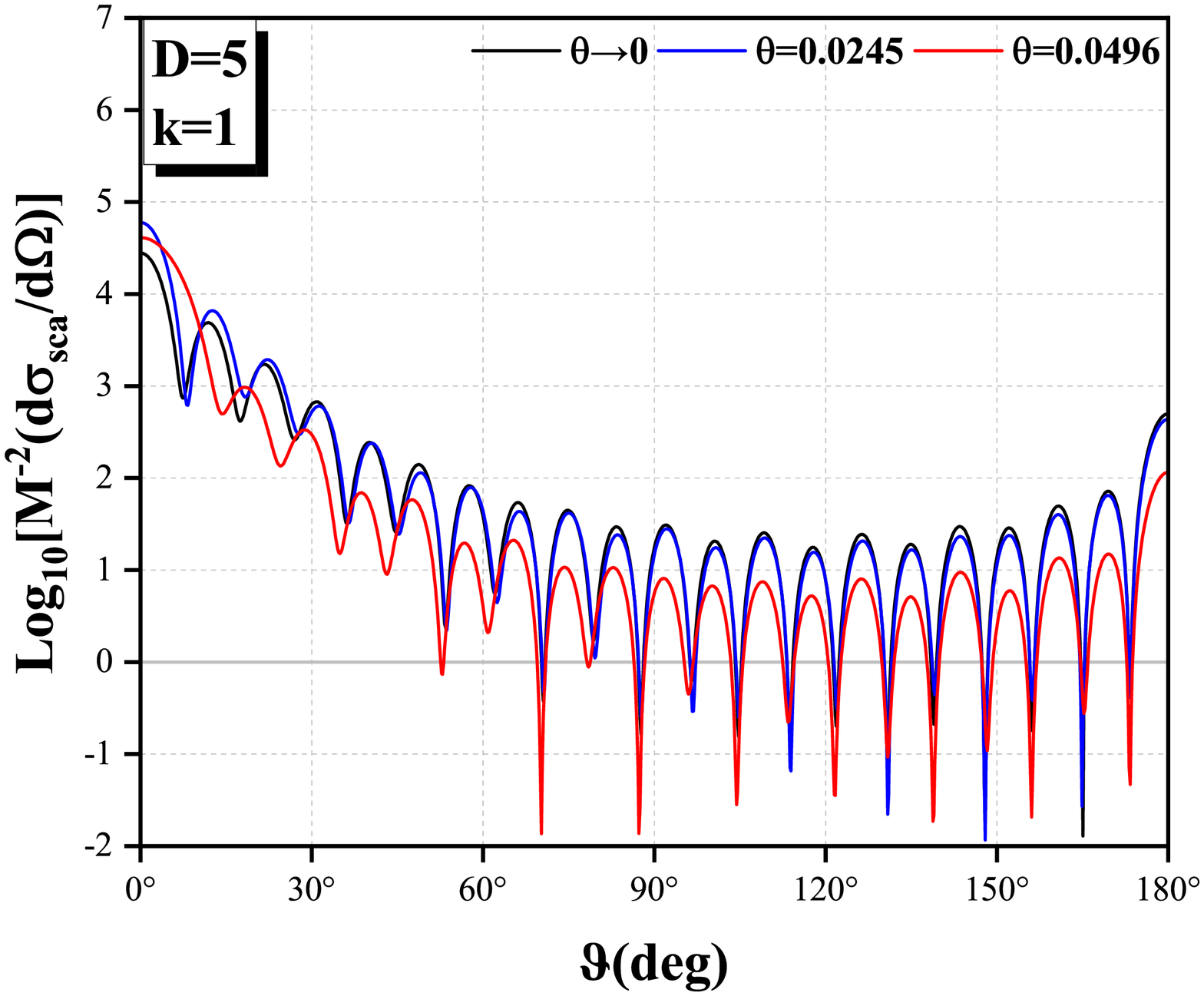}
\includegraphics[width=0.3\textwidth]{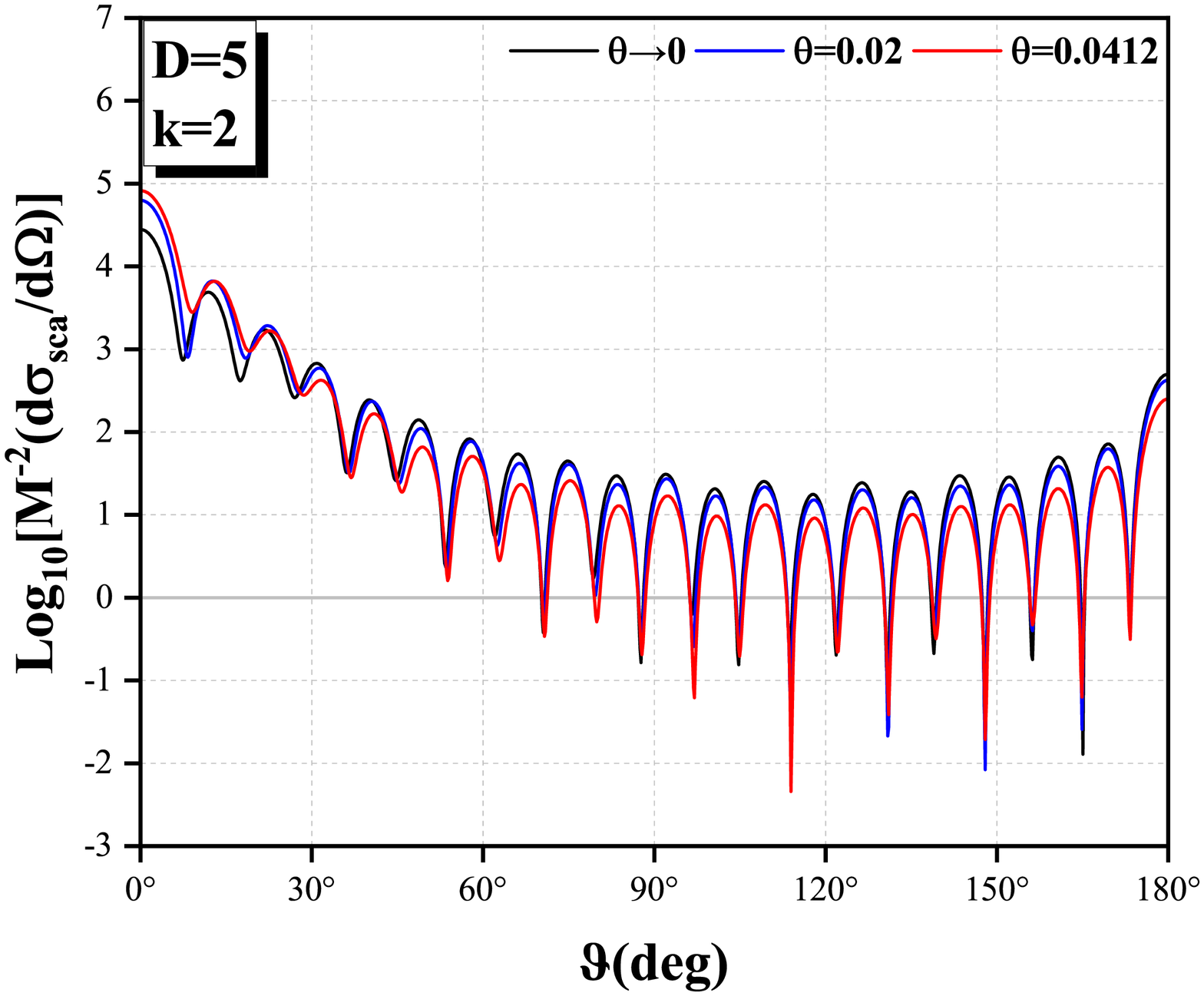}
\includegraphics[width=0.3\textwidth]{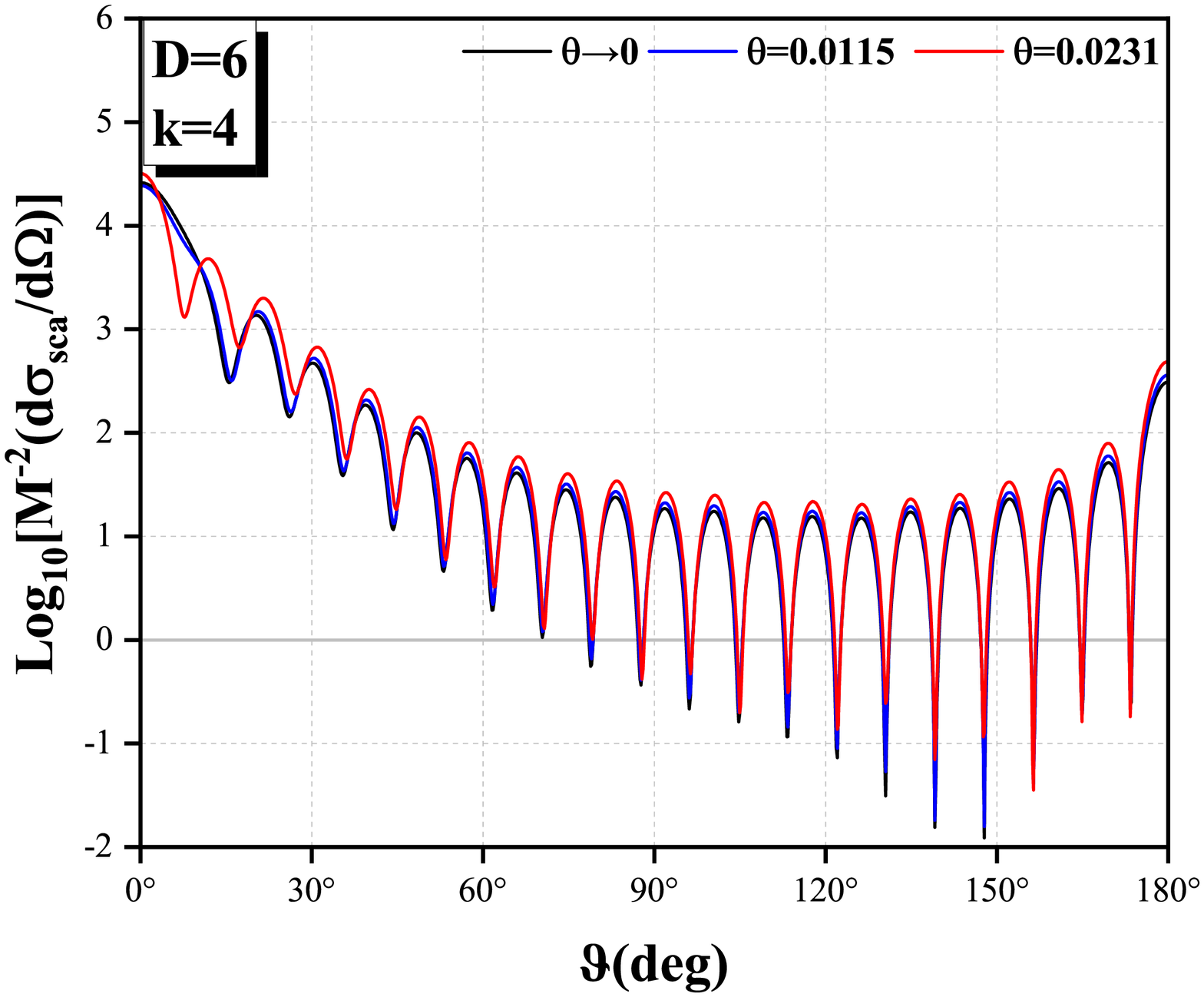}
\includegraphics[width=0.3\textwidth]{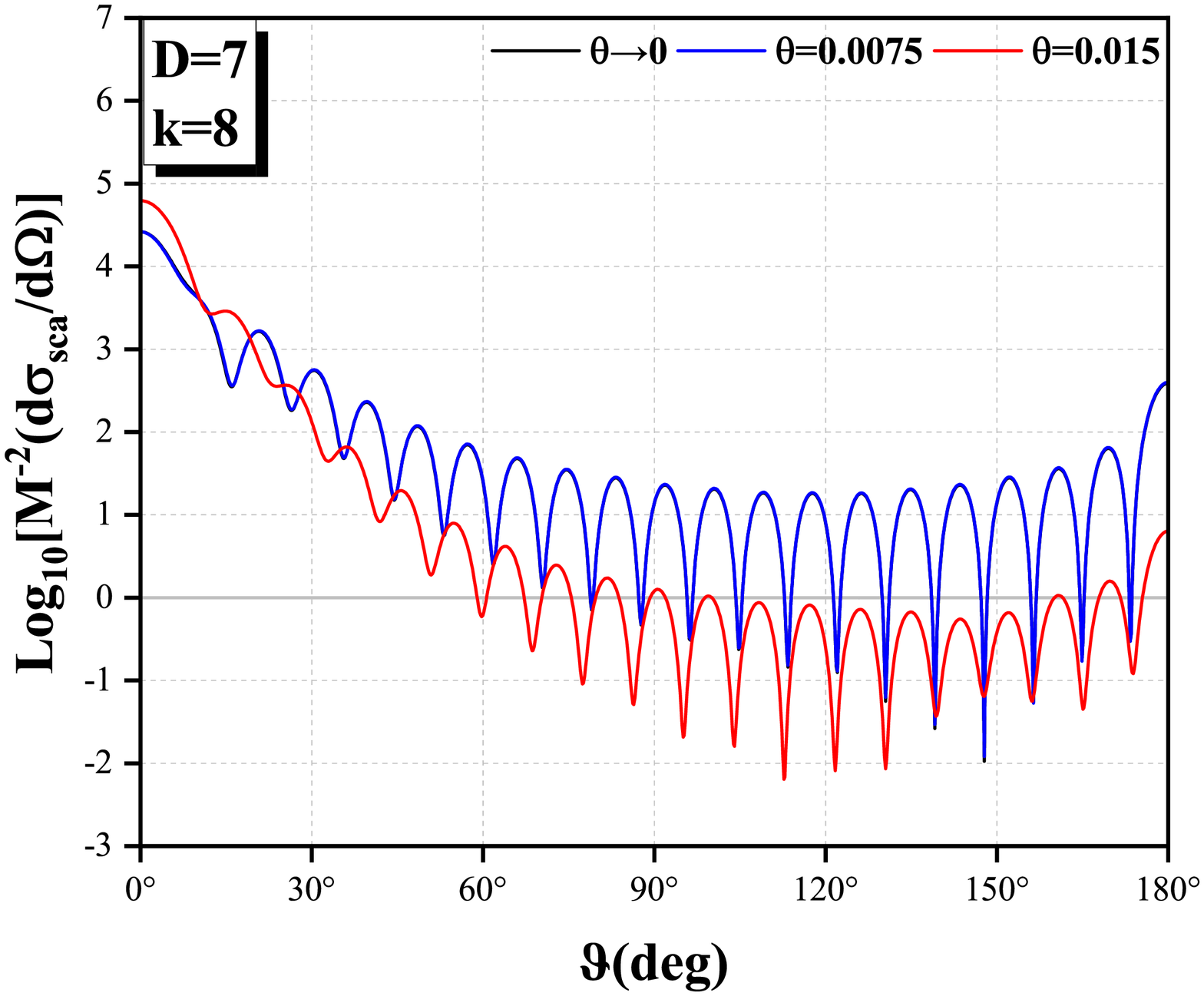}
\includegraphics[width=0.3\textwidth]{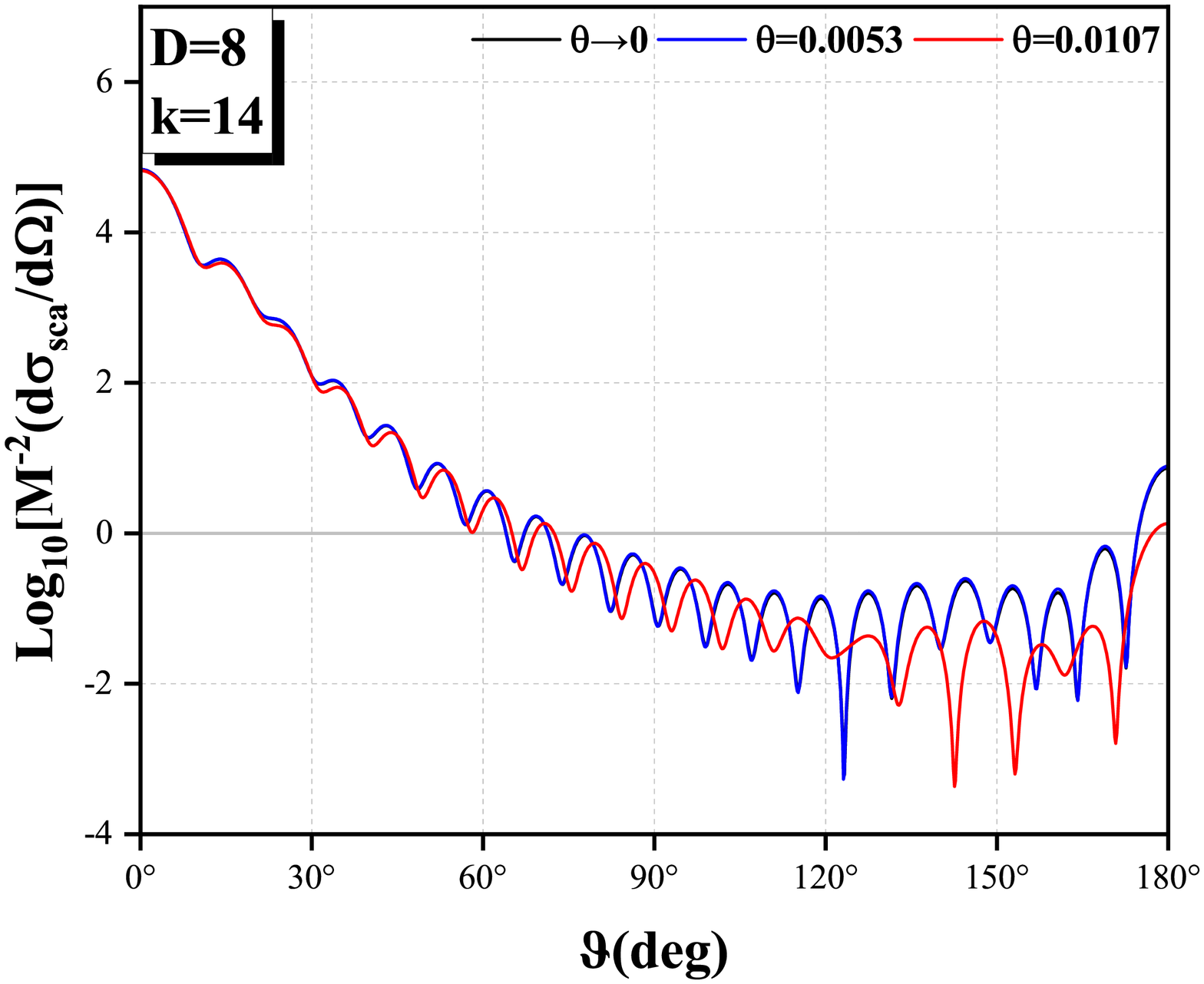}
\includegraphics[width=0.3\textwidth]{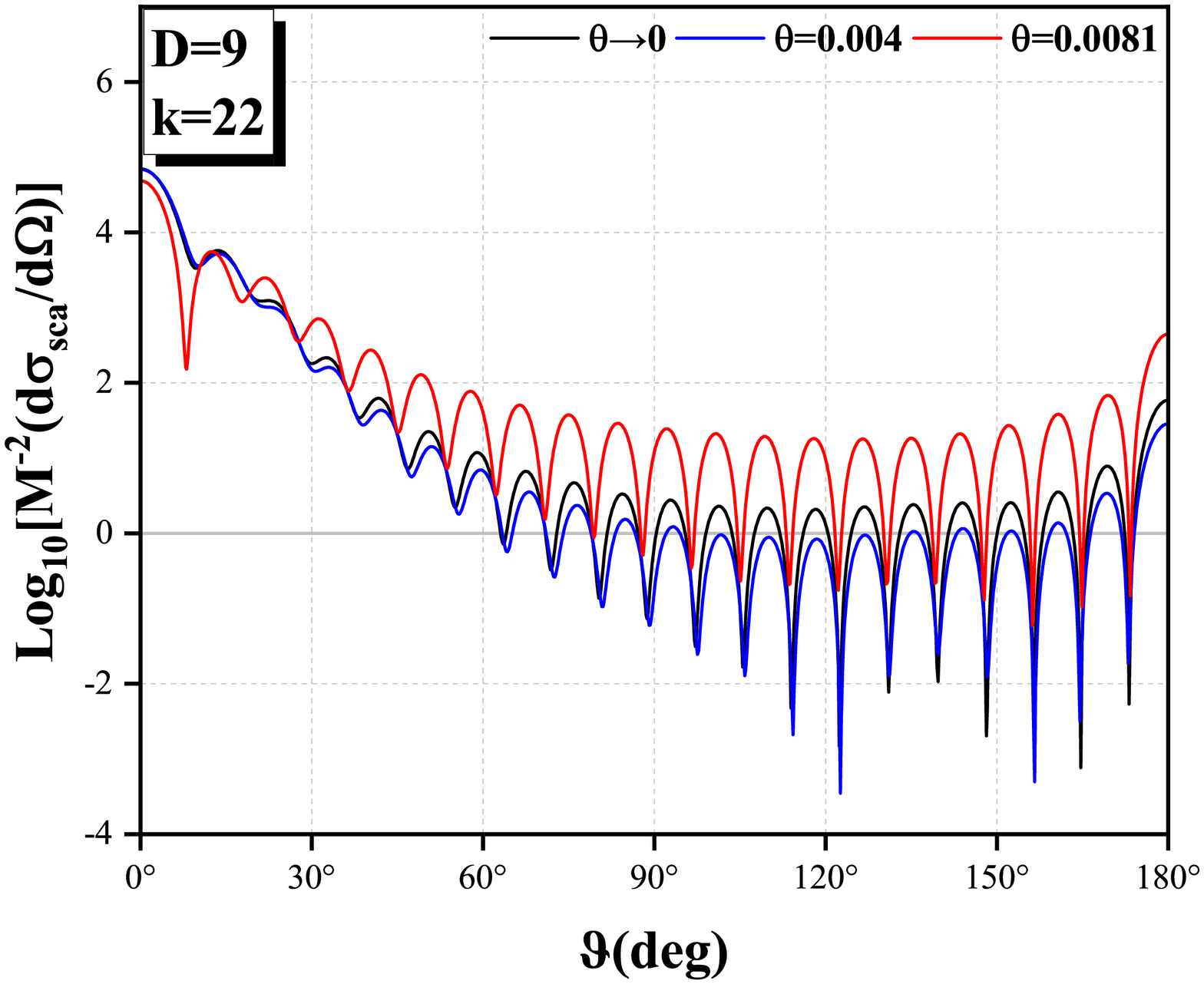}
\includegraphics[width=0.3\textwidth]{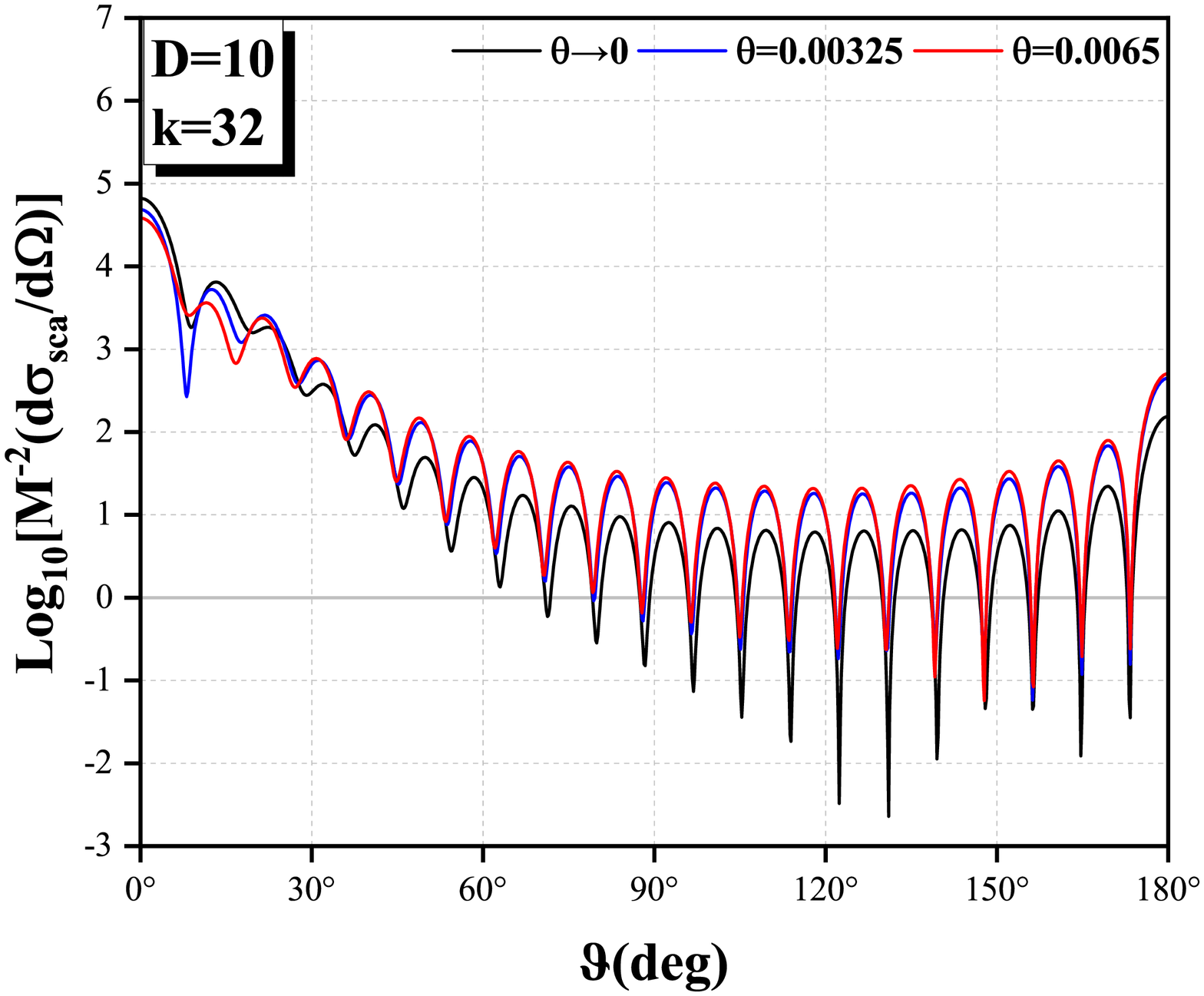}
\includegraphics[width=0.3\textwidth]{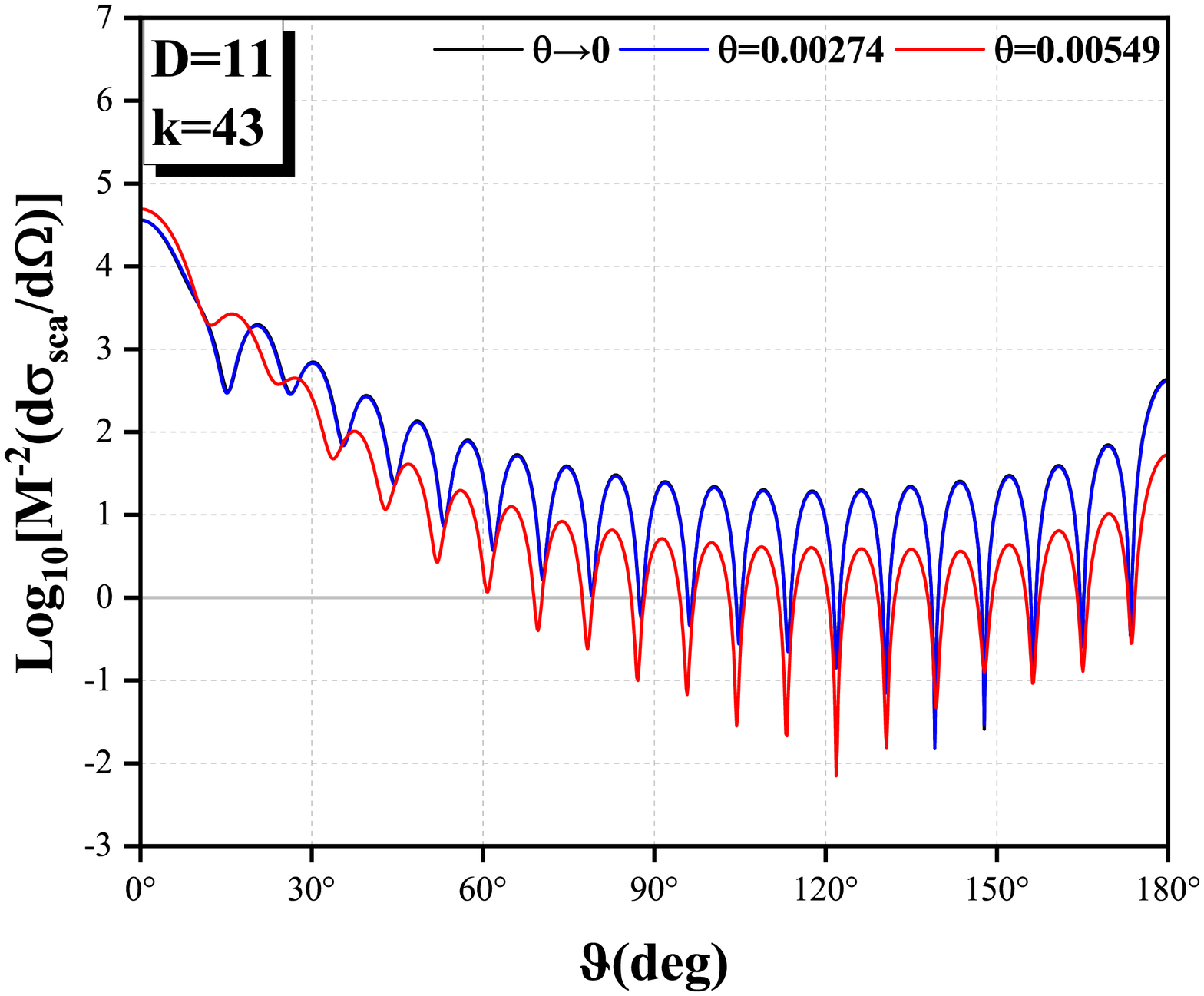}
\caption{\label{fig4}The logarithm values of the differential scattering cross sections of high dimensional non\textendash{}commutative black holes with different values of $k$ and $D$ are plotted in the range of $\vartheta=0^{\circ}$ to $\vartheta=180^{\circ}$, 
with $\omega M$=1 fixed.
}
\end{figure*}

\section{Numerical results} \label{sec3}

To determine the absorption and scattering cross sections of the scalar field dispersed by the black hole in this study, 
we must solve the radial equation with appropriate boundary conditions. 
For our scattering problem, 
we consider plane scalar waves incoming from the infinite null past. 
Therefore, 
we are interested in solutions of Eq. (\ref{eq14}) subjected to the following boundary conditions.
\begin{equation} \label{eq26}
\phi_{\omega l}(r_{*})\approx
\begin{cases}
A_{\text{in}}e^{-i \omega r_{*}}+A_{\text{out}}e^{i \omega r_{*}} &\text{for } r_{*}\to\infty;\\
e^{-i \omega r_{*}} &\text{for } r_{*}\to-\infty
\end{cases}
\end{equation}
and satisfy the conservation relation $1+|A_{\text{out}}|^{2}=|A_{\text{in}}|^{2}$. 
The scattered wave's phase shift is defined as
\begin{equation} \label{eq27}
e^{i\omega\delta_{l}}=(-1)^{l+1}\frac{A_{\text{out}}}{A_{\text{in}}}.
\end{equation}
According to quantum mechanics, 
when an incident wave travels through a potential barrier, 
some of it is reflected back and part of it is transmitted across the potential barrier. 
The transmission amplitude $T_{l}$ and reflection amplitude $R_{l}$ fulfill flux conservation.
\begin{equation} \label{eq28}
\lvert T_{l}\rvert^{2}+\vert R_{l}\rvert^{2}=1,
\end{equation}
where $T_{l}$ and $R_{l}$ denote the transmission amplitude $T_{l}=1/A_{\text{in}}$ and the reflection amplitude $R_{l}=A_{\text{out}}/A_{\text{in}}$, respectively.

\subsection{Absorption cross section} \label{subsec3.1}

It is well known that the total absorption cross section can be calculated as
\begin{equation} \label{eq29}
\sigma_{\text{abs}}=\sum\limits_{l=0}^{\infty}\sigma^{(l)}_{\text{abs}},
\end{equation}
where $\sigma^{(l)}_{\text{abs}}$ denotes the partial absorption cross section and can be expressed by transmission coefficient:
\begin{equation} \label{eq30}
\sigma^{(l)}_{\text{abs}}=\frac{\pi}{\omega^{2}}(2l+1)(1-\lvert e^{2i\delta_{l}}\rvert^{2})=\frac{\pi}{\omega^{2}}(2l+1)\lvert T_{\omega l}\rvert^{2}.
\end{equation}

In FIG. \ref{fig3}, 
For various values of the $k$ and $D$ parameters, 
we illustrate the partial absorption cross sections $\sigma^{(l)}_{\text{abs}}$ corresponding to high dimensional non\textendash{}commutative black holes. 
i.e. $l=0, 1, 2, 3, 4, 5$ and total absorption cross sections $\sigma_{\text{abs}}$ of the $D$\textendash{}dimensional non\textendash{}commutative black holes with different metric parameters. 
The mass $M$ is normalized to 1. 
It is obvious that the partial absorption cross section $\sigma^{(l)}_{\text{abs}}$ tends to vanish as $\omega M$ increases. 
In the null energy limit, 
the partial wave with $l=0$ provides a nonzero absorption cross section \cite{bib47}. 
Furthermore, 
for each value of $l>0$, 
the corresponding partial absorption cross section $\sigma^{(l)}_{\text{abs}}$ starts from zero, 
reaches a peak, 
and decreases asymptotically in the high frequency zone. 
The larger the value of $l$ is, 
(i)the smaller the corresponding maximum of $\sigma^{(l)}_{\text{abs}}$ is and (ii) the larger the value of $\omega M$ associated with the peak of $\sigma^{(l)}_{\text{abs}}$ is. 
This is compatible with the fact that the larger the value of $l$ is, 
the higher the scattering potential $V_{\theta,k}$ is.

FIG. \ref{fig3} further demonstrates that the parameter $\theta$ has a slight influence on the absorption cross section.
For $D=4$ cases with $\theta\to\theta_{\text{max}}$, 
the absorption cross section shows a increasing trend. 
But in $D\ge5$ cases,
the absorption cross section shows a decreasing trend. 
The total absorption cross section $\sigma_{\text{abs}}$ oscillates approaching the geometric optics limit. 
The total absorption cross section goes to the geometrical optical limit $\sigma^{\text{hf}}_{\text{abs}}=\pi b_{\text{cr}}^{2}$ \cite{bib47} when $\omega M\gg1$ reaches the high frequency zone, 
where $b_{\text{cr}}$ is the crucial impact parameter \cite{bib63}. 
The larger dimensionless number $\omega M$ is required to attain the geometrical optical limit as the $D$ increases.

For $D=4$,
We show the situation of $k=$0, 1 and 2, respectively.
With the increase of $k$,
$\theta_{\text{max}}$ is getting smaller and smaller.
We can see that the $k$ value has little effect on the absorption cross section.
The absorption cross section tends to the Schwarzschild case when $\theta\to0$ in the case of $D=4$ and $k=0$, 
which is consistent with the predicted outcome. 

When $D\ge5$, 
the situation changes dramatically. 
As $D$ climbs, 
the absorption cross section decreases as usually. 
However, 
in the low frequency range, 
when $\theta$ is tiny and $l$ is relatively small, 
the absorption cross section begins to fluctuate and has a significant peak. 
The absorption cross section curve in the low frequency zone is smooth when $\theta$ is greater and goes to $\theta_{\text{max}}$. 
In both circumstances, 
the high frequency area is smooth and the oscillations disappear. 
In contrast to the situation of $D=4$ dimension, 
the total absorption cross section oscillates slightly and approaches the limit value at the much larger $\omega M$.

\subsection{Scattering cross section} \label{subsec3.2}

It is widely known that the scattering amplitude is defined as
\begin{equation} \label{eq31}
g(\vartheta)=\frac{1}{2i\omega}\sum\limits_{l=0}^{\infty}(2l+1)(e^{2i\delta_{l}}-1)P_{l}(\cos\vartheta)
\end{equation}
and the differential scattering cross section
\begin{equation} \label{eq32}
\frac{\mathrm{d}\sigma_{\text{sca}}}{\mathrm{d}\Omega}=\lvert g(\vartheta)\rvert^{2}.
\end{equation}
As a direct consequence, 
the scattering cross section is obtained
\begin{equation} \label{eq33}
\sigma_{\text{sca}}(\omega)=\int\frac{\mathrm{d}\sigma}{\mathrm{d}\Omega}\mathrm{d}\Omega=\frac{1}{2i\omega}\sum\limits_{l=0}^{\infty}(2l+1)\lvert e^{2i\delta_{l}}-1\rvert^{2}.
\end{equation}

FIG. \ref{fig4} shows the differential scattering cross section as a function of the scattering angle at $\omega M$=1. 
For some black holes, 
we find the scattering flux is strengthened and its width becomes narrower in the forward direction. 
The scalar field scattering becomes more concentrated. 
With fixed frequency in the low frequency range, 
the glory peak is higher and the glory width becomes narrower. 
As a result, 
the glory phenomena at the forward and backward lends itself advantageously to astronomical observation. 
Furthermore, 
we can see that the impact of non\textendash{}commutative on the differential scattering cross section is stronger at large angles.

In the cases of $D=4$,
the glory peaks becomes narrower as the non\textendash{}commutative $\theta$ increases.
However,
compared with the high\textendash{}dimensional cases,
the glory peaks are wider than high\textendash{}dimensional cases.
When $D\ge5$,
we also fixed the $\omega M=1$,
but the scattering pattern is similar to high frequency scattering.
The glory peaks is quite narrow.

\section{Conclusions} \label{sec4}

In summary, 
we have investigated the scattering and absorption cross section of the massless scalar field from several high dimensional non\textendash{}commutative black hole with smeared matter sources. 
We investigate that the interaction between frequency $\omega M$ and parameters $D$, $k$, $\theta$, respectively.
We exhibit the results of these parameters with varying values.

Let us summarize the scattering and absorption of massless scalar fields by high dimensional non\textendash{}commutative black hole.
For absorption cross section in $D=4$ with fixed $k$,
our computational results indicated that the larger $\theta$ is, 
the lower the associated total absorption cross section is. 
The total absorption cross section oscillates approaches the geometrical optical limit in the high frequency zone. 
But in $D\ge5$ cases,
the absorption cross section shows a decreasing trend.
The total absorption cross section oscillates approaches the geometrical optical limit in the much higher frequency region compare with the $D=4$ cases and the total absorption cross section oscillates slightly.

While for the scattering cross section at fixed $\omega M$,
the scattering flux is intensified and its width narrows in the forward direction for some black holes. 
The scattering of the scalar field gets more intense and the forward and backward directions of glory phenomenon lends itself well to astronomical observation.
The glory peaks get thinner as the non-commutative $\theta$ rises in the circumstances when $D = 4$.
When $D \ge5$, 
we likewise set $\omega M = 1$, 
but the scattering pattern resembles high frequency scattering and the glory width are relatively small comparewith $D=4$.
The influence of non\textendash{}commutative on the differential scattering cross section is particularly pronounced at large angles.

In theory, high-precision measurements of massless scalar wave fluxes scattered by black holes might one day be used to estimate black hole's hairs.
A more immediate possibility is that scattering and absorption patterns may be observed with black hole analogue systems created in the laboratory \cite{bib70,bib71}.

\begin{acknowledgments}

The authors thank Dr. Z.N. Yan for his positive help and useful discussion. 
This work was supported by the National Key Research and Develop Program of China under Contract No. 2018YFA0404404.

\end{acknowledgments}

\bibliography{HDSBHwithScaler}

\begin{thebibliography}{71}%
\makeatletter
\providecommand \@ifxundefined [1]{%
 \@ifx{#1\undefined}
}%
\providecommand \@ifnum [1]{%
 \ifnum #1\expandafter \@firstoftwo
 \else \expandafter \@secondoftwo
 \fi
}%
\providecommand \@ifx [1]{%
 \ifx #1\expandafter \@firstoftwo
 \else \expandafter \@secondoftwo
 \fi
}%
\providecommand \natexlab [1]{#1}%
\providecommand \enquote  [1]{``#1''}%
\providecommand \bibnamefont  [1]{#1}%
\providecommand \bibfnamefont [1]{#1}%
\providecommand \citenamefont [1]{#1}%
\providecommand \href@noop [0]{\@secondoftwo}%
\providecommand \href [0]{\begingroup \@sanitize@url \@href}%
\providecommand \@href[1]{\@@startlink{#1}\@@href}%
\providecommand \@@href[1]{\endgroup#1\@@endlink}%
\providecommand \@sanitize@url [0]{\catcode `\\12\catcode `\$12\catcode
  `\&12\catcode `\#12\catcode `\^12\catcode `\_12\catcode `\%12\relax}%
\providecommand \@@startlink[1]{}%
\providecommand \@@endlink[0]{}%
\providecommand \url  [0]{\begingroup\@sanitize@url \@url }%
\providecommand \@url [1]{\endgroup\@href {#1}{\urlprefix }}%
\providecommand \urlprefix  [0]{URL }%
\providecommand \Eprint [0]{\href }%
\providecommand \doibase [0]{https://doi.org/}%
\providecommand \selectlanguage [0]{\@gobble}%
\providecommand \bibinfo  [0]{\@secondoftwo}%
\providecommand \bibfield  [0]{\@secondoftwo}%
\providecommand \translation [1]{[#1]}%
\providecommand \BibitemOpen [0]{}%
\providecommand \bibitemStop [0]{}%
\providecommand \bibitemNoStop [0]{.\EOS\space}%
\providecommand \EOS [0]{\spacefactor3000\relax}%
\providecommand \BibitemShut  [1]{\csname bibitem#1\endcsname}%
\let\auto@bib@innerbib\@empty
\bibitem [{\citenamefont {Nicolini}(2009)}]{bib1}%
  \BibitemOpen
  \bibfield  {author} {\bibinfo {author} {\bibfnamefont {P.}~\bibnamefont
  {Nicolini}},\ }\bibfield  {title} {\bibinfo {title} {Noncommutative black
  holes, the final appeal to quantum gravity: A review},\ }\href
  {https://doi.org/10.1142/S0217751X09043353} {\bibfield  {journal} {\bibinfo
  {journal} {Int.J.Mod.Phys.A}\ }\textbf {\bibinfo {volume} {24}},\ \bibinfo
  {pages} {1229} (\bibinfo {year} {2009})}\BibitemShut {NoStop}%
\bibitem [{\citenamefont {Szabo}(2006)}]{bib2}%
  \BibitemOpen
  \bibfield  {author} {\bibinfo {author} {\bibfnamefont {R.~J.}\ \bibnamefont
  {Szabo}},\ }\bibfield  {title} {\bibinfo {title} {Symmetry, gravity and
  noncommutativity},\ }\href {https://doi.org/10.1088/0264-9381/23/22/r01}
  {\bibfield  {journal} {\bibinfo  {journal} {Class.Quant.Grav.}\ }\textbf
  {\bibinfo {volume} {23}},\ \bibinfo {pages} {R199} (\bibinfo {year}
  {2006})}\BibitemShut {NoStop}%
\bibitem [{\citenamefont {Snyder}(1947)}]{bib3}%
  \BibitemOpen
  \bibfield  {author} {\bibinfo {author} {\bibfnamefont {H.~S.}\ \bibnamefont
  {Snyder}},\ }\bibfield  {title} {\bibinfo {title} {Quantized
  space\textendash{}time},\ }\href {https://doi.org/10.1103/PhysRev.71.38}
  {\bibfield  {journal} {\bibinfo  {journal} {Phys.Rev.}\ }\textbf {\bibinfo
  {volume} {71}},\ \bibinfo {pages} {38} (\bibinfo {year} {1947})}\BibitemShut
  {NoStop}%
\bibitem [{\citenamefont {Smailagic}\ and\ \citenamefont
  {Spallucci}(2003{\natexlab{a}})}]{bib4}%
  \BibitemOpen
  \bibfield  {author} {\bibinfo {author} {\bibfnamefont {A.}~\bibnamefont
  {Smailagic}}\ and\ \bibinfo {author} {\bibfnamefont {E.}~\bibnamefont
  {Spallucci}},\ }\bibfield  {title} {\bibinfo {title} {Feynman path integral
  on the noncommutative plane},\ }\href
  {https://doi.org/10.1088/0305-4470/36/33/101} {\bibfield  {journal} {\bibinfo
   {journal} {J. Phys. A}\ }\textbf {\bibinfo {volume} {36}},\ \bibinfo {pages}
  {L467} (\bibinfo {year} {2003}{\natexlab{a}})}\BibitemShut {NoStop}%
\bibitem [{\citenamefont {Smailagic}\ and\ \citenamefont
  {Spallucci}(2003{\natexlab{b}})}]{bib5}%
  \BibitemOpen
  \bibfield  {author} {\bibinfo {author} {\bibfnamefont {A.}~\bibnamefont
  {Smailagic}}\ and\ \bibinfo {author} {\bibfnamefont {E.}~\bibnamefont
  {Spallucci}},\ }\bibfield  {title} {\bibinfo {title} {Uv divergence free qft
  on noncommutative plane},\ }\href
  {https://doi.org/10.1088/0305-4470/36/39/103} {\bibfield  {journal} {\bibinfo
   {journal} {J. Phys. A}\ }\textbf {\bibinfo {volume} {36}},\ \bibinfo {pages}
  {L517} (\bibinfo {year} {2003}{\natexlab{b}})}\BibitemShut {NoStop}%
\bibitem [{\citenamefont {Susskind}(1993)}]{bib6}%
  \BibitemOpen
  \bibfield  {author} {\bibinfo {author} {\bibfnamefont {L.}~\bibnamefont
  {Susskind}},\ }\bibfield  {title} {\bibinfo {title} {String theory and the
  principles of black hole complementarity},\ }\href
  {https://doi.org/10.1103/PhysRevLett.71.2367} {\bibfield  {journal} {\bibinfo
   {journal} {Phys. Rev. Lett.}\ }\textbf {\bibinfo {volume} {71}},\ \bibinfo
  {pages} {2367} (\bibinfo {year} {1993})}\BibitemShut {NoStop}%
\bibitem [{\citenamefont {Nicolini}\ \emph {et~al.}(2006)\citenamefont
  {Nicolini}, \citenamefont {Smailagic},\ and\ \citenamefont
  {Spallucci}}]{bib7}%
  \BibitemOpen
  \bibfield  {author} {\bibinfo {author} {\bibfnamefont {P.}~\bibnamefont
  {Nicolini}}, \bibinfo {author} {\bibfnamefont {A.}~\bibnamefont
  {Smailagic}},\ and\ \bibinfo {author} {\bibfnamefont {E.}~\bibnamefont
  {Spallucci}},\ }\bibfield  {title} {\bibinfo {title} {Noncommutative geometry
  inspired schwarzschild black hole},\ }\href
  {https://doi.org/10.1016/j.physletb.2005.11.004} {\bibfield  {journal}
  {\bibinfo  {journal} {Phys. Lett. B}\ }\textbf {\bibinfo {volume} {632}},\
  \bibinfo {pages} {547} (\bibinfo {year} {2006})}\BibitemShut {NoStop}%
\bibitem [{\citenamefont {Seiberg}\ and\ \citenamefont {Witten}(1999)}]{bib8}%
  \BibitemOpen
  \bibfield  {author} {\bibinfo {author} {\bibfnamefont {N.}~\bibnamefont
  {Seiberg}}\ and\ \bibinfo {author} {\bibfnamefont {E.}~\bibnamefont
  {Witten}},\ }\bibfield  {title} {\bibinfo {title} {String theory and
  noncommutative geometry},\ }\href
  {https://doi.org/10.1088/1126-6708/1999/09/032} {\bibfield  {journal}
  {\bibinfo  {journal} {JHEP}\ }\textbf {\bibinfo {volume} {1999}}\bibinfo
  {number} { (09)},\ \bibinfo {pages} {032}}\BibitemShut {NoStop}%
\bibitem [{\citenamefont {Nozari}\ and\ \citenamefont
  {Mehdipour}(2008)}]{bib9}%
  \BibitemOpen
\bibfield  {number} {  }\bibfield  {author} {\bibinfo {author} {\bibfnamefont
  {K.}~\bibnamefont {Nozari}}\ and\ \bibinfo {author} {\bibfnamefont {S.~H.}\
  \bibnamefont {Mehdipour}},\ }\bibfield  {title} {\bibinfo {title} {Hawking
  radiation as quantum tunneling from a noncommutative schwarzschild black
  hole},\ }\href {https://doi.org/10.1088/0264-9381/25/17/175015} {\bibfield
  {journal} {\bibinfo  {journal} {Class. Quant. Grav.}\ }\textbf {\bibinfo
  {volume} {25}},\ \bibinfo {pages} {175015} (\bibinfo {year}
  {2008})}\BibitemShut {NoStop}%
\bibitem [{\citenamefont {Myung}\ and\ \citenamefont {Yoon}(2009)}]{bib10}%
  \BibitemOpen
  \bibfield  {author} {\bibinfo {author} {\bibfnamefont {Y.~S.}\ \bibnamefont
  {Myung}}\ and\ \bibinfo {author} {\bibfnamefont {M.}~\bibnamefont {Yoon}},\
  }\bibfield  {title} {\bibinfo {title} {Regular black hole in three
  dimensions},\ }\href {https://doi.org/10.1140/epjc/s10052-009-1036-9}
  {\bibfield  {journal} {\bibinfo  {journal} {Eur. Phys. J. C}\ }\textbf
  {\bibinfo {volume} {62}},\ \bibinfo {pages} {405} (\bibinfo {year}
  {2009})}\BibitemShut {NoStop}%
\bibitem [{\citenamefont {Ansoldi}\ \emph {et~al.}(2007)\citenamefont
  {Ansoldi}, \citenamefont {Nicolini}, \citenamefont {Smailagic},\ and\
  \citenamefont {Spallucci}}]{bib11}%
  \BibitemOpen
  \bibfield  {author} {\bibinfo {author} {\bibfnamefont {S.}~\bibnamefont
  {Ansoldi}}, \bibinfo {author} {\bibfnamefont {P.}~\bibnamefont {Nicolini}},
  \bibinfo {author} {\bibfnamefont {A.}~\bibnamefont {Smailagic}},\ and\
  \bibinfo {author} {\bibfnamefont {E.}~\bibnamefont {Spallucci}},\ }\bibfield
  {title} {\bibinfo {title} {Noncommutative geometry inspired charged black
  holes},\ }\href {https://doi.org/10.1016/j.physletb.2006.12.020} {\bibfield
  {journal} {\bibinfo  {journal} {Phys. Lett. B}\ }\textbf {\bibinfo {volume}
  {645}},\ \bibinfo {pages} {261} (\bibinfo {year} {2007})}\BibitemShut
  {NoStop}%
\bibitem [{\citenamefont {Nicolini}\ and\ \citenamefont
  {Torrieri}(2011)}]{bib12}%
  \BibitemOpen
  \bibfield  {author} {\bibinfo {author} {\bibfnamefont {P.}~\bibnamefont
  {Nicolini}}\ and\ \bibinfo {author} {\bibfnamefont {G.}~\bibnamefont
  {Torrieri}},\ }\bibfield  {title} {\bibinfo {title} {The
  hawking\textendash{}page crossover in noncommutative
  anti\textendash{}desitter space},\ }\href
  {https://doi.org/10.1007/JHEP08(2011)097} {\bibfield  {journal} {\bibinfo
  {journal} {JHEP}\ }\textbf {\bibinfo {volume} {2011}}\bibinfo  {number} {
  (08)},\ \bibinfo {pages} {097}}\BibitemShut {NoStop}%
\bibitem [{\citenamefont {Modesto}\ and\ \citenamefont
  {Nicolini}(2010)}]{bib13}%
  \BibitemOpen
\bibfield  {number} {  }\bibfield  {author} {\bibinfo {author} {\bibfnamefont
  {L.}~\bibnamefont {Modesto}}\ and\ \bibinfo {author} {\bibfnamefont
  {P.}~\bibnamefont {Nicolini}},\ }\bibfield  {title} {\bibinfo {title}
  {Charged rotating noncommutative black holes},\ }\href
  {https://doi.org/10.1103/PhysRevD.82.104035} {\bibfield  {journal} {\bibinfo
  {journal} {Phys. Rev. D}\ }\textbf {\bibinfo {volume} {82}},\ \bibinfo
  {pages} {104035} (\bibinfo {year} {2010})}\BibitemShut {NoStop}%
\bibitem [{\citenamefont {Nozari}\ and\ \citenamefont
  {Mehdipour}(2009)}]{bib14}%
  \BibitemOpen
  \bibfield  {author} {\bibinfo {author} {\bibfnamefont {K.}~\bibnamefont
  {Nozari}}\ and\ \bibinfo {author} {\bibfnamefont {S.~H.}\ \bibnamefont
  {Mehdipour}},\ }\bibfield  {title} {\bibinfo {title}
  {Parikh\textendash{}wilczek tunneling from noncommutative higher dimensional
  black holes},\ }\href {https://doi.org/10.1088/1126-6708/2009/03/061}
  {\bibfield  {journal} {\bibinfo  {journal} {JHEP}\ }\textbf {\bibinfo
  {volume} {2009}}\bibinfo  {number} { (03)},\ \bibinfo {pages}
  {061}}\BibitemShut {NoStop}%
\bibitem [{\citenamefont {Rizzo}(2006)}]{bib15}%
  \BibitemOpen
\bibfield  {number} {  }\bibfield  {author} {\bibinfo {author} {\bibfnamefont
  {T.~G.}\ \bibnamefont {Rizzo}},\ }\bibfield  {title} {\bibinfo {title}
  {Noncommutative inspired black holes in extra dimensions},\ }\href
  {https://doi.org/10.1088/1126-6708/2006/09/021} {\bibfield  {journal}
  {\bibinfo  {journal} {JHEP}\ }\textbf {\bibinfo {volume} {2006}}\bibinfo
  {number} { (09)},\ \bibinfo {pages} {021}}\BibitemShut {NoStop}%
\bibitem [{\citenamefont {Spallucci}\ \emph {et~al.}(2009)\citenamefont
  {Spallucci}, \citenamefont {Smailagic},\ and\ \citenamefont
  {Nicolini}}]{bib16}%
  \BibitemOpen
\bibfield  {number} {  }\bibfield  {author} {\bibinfo {author} {\bibfnamefont
  {E.}~\bibnamefont {Spallucci}}, \bibinfo {author} {\bibfnamefont
  {A.}~\bibnamefont {Smailagic}},\ and\ \bibinfo {author} {\bibfnamefont
  {P.}~\bibnamefont {Nicolini}},\ }\bibfield  {title} {\bibinfo {title}
  {Non\textendash{}commutative geometry inspired higher\textendash{}dimensional
  charged black holes},\ }\href
  {https://doi.org/10.1016/j.physletb.2008.11.030} {\bibfield  {journal}
  {\bibinfo  {journal} {Phys. Lett. B}\ }\textbf {\bibinfo {volume} {670}},\
  \bibinfo {pages} {449} (\bibinfo {year} {2009})}\BibitemShut {NoStop}%
\bibitem [{\citenamefont {Nozari}\ and\ \citenamefont
  {Mehdipour}(2010)}]{bib17}%
  \BibitemOpen
  \bibfield  {author} {\bibinfo {author} {\bibfnamefont {K.}~\bibnamefont
  {Nozari}}\ and\ \bibinfo {author} {\bibfnamefont {S.~H.}\ \bibnamefont
  {Mehdipour}},\ }\bibfield  {title} {\bibinfo {title} {Noncommutative inspired
  reissner\textendash{}nordstr{\"o}m black holes in large extra dimensions},\
  }\href {https://doi.org/10.1088/0253-6102/53/3/20} {\bibfield  {journal}
  {\bibinfo  {journal} {Commun. Theor. Phys.}\ }\textbf {\bibinfo {volume}
  {53}},\ \bibinfo {pages} {503} (\bibinfo {year} {2010})}\BibitemShut
  {NoStop}%
\bibitem [{\citenamefont {Banerjee}\ \emph {et~al.}(2008)\citenamefont
  {Banerjee}, \citenamefont {Majhi},\ and\ \citenamefont {Samanta}}]{bib18}%
  \BibitemOpen
  \bibfield  {author} {\bibinfo {author} {\bibfnamefont {R.}~\bibnamefont
  {Banerjee}}, \bibinfo {author} {\bibfnamefont {B.~R.}\ \bibnamefont
  {Majhi}},\ and\ \bibinfo {author} {\bibfnamefont {S.}~\bibnamefont
  {Samanta}},\ }\bibfield  {title} {\bibinfo {title} {Noncommutative black hole
  thermodynamics},\ }\href {https://doi.org/10.1103/PhysRevD.77.124035}
  {\bibfield  {journal} {\bibinfo  {journal} {Phys. Rev. D}\ }\textbf {\bibinfo
  {volume} {77}},\ \bibinfo {pages} {124035} (\bibinfo {year}
  {2008})}\BibitemShut {NoStop}%
\bibitem [{\citenamefont {Myung}\ \emph {et~al.}(2007)\citenamefont {Myung},
  \citenamefont {Kim},\ and\ \citenamefont {Park}}]{bib19}%
  \BibitemOpen
  \bibfield  {author} {\bibinfo {author} {\bibfnamefont {Y.~S.}\ \bibnamefont
  {Myung}}, \bibinfo {author} {\bibfnamefont {Y.}~\bibnamefont {Kim}},\ and\
  \bibinfo {author} {\bibfnamefont {Y.}~\bibnamefont {Park}},\ }\bibfield
  {title} {\bibinfo {title} {Thermodynamics and evaporation of the
  noncommutative black hole},\ }\href
  {https://doi.org/10.1088/1126-6708/2007/02/012} {\bibfield  {journal}
  {\bibinfo  {journal} {JHEP}\ }\textbf {\bibinfo {volume} {2007}}\bibinfo
  {number} { (22)},\ \bibinfo {pages} {012}}\BibitemShut {NoStop}%
\bibitem [{\citenamefont {Nozari}\ and\ \citenamefont
  {Fazlpour}(2008)}]{bib20}%
  \BibitemOpen
\bibfield  {number} {  }\bibfield  {author} {\bibinfo {author} {\bibfnamefont
  {K.}~\bibnamefont {Nozari}}\ and\ \bibinfo {author} {\bibfnamefont
  {B.}~\bibnamefont {Fazlpour}},\ }\bibfield  {title} {\bibinfo {title}
  {Reissner\textendash{}nordstr{\"o}m black hole thermodynamics in
  noncommutative spaces},\ }\href@noop {} {\bibfield  {journal} {\bibinfo
  {journal} {Acta Phys. Polon. B}\ }\textbf {\bibinfo {volume} {39}},\ \bibinfo
  {pages} {1363} (\bibinfo {year} {2008})}\BibitemShut {NoStop}%
\bibitem [{\citenamefont {Miao}\ and\ \citenamefont {Xu}(2016)}]{bib21}%
  \BibitemOpen
  \bibfield  {author} {\bibinfo {author} {\bibfnamefont {Y.}~\bibnamefont
  {Miao}}\ and\ \bibinfo {author} {\bibfnamefont {Z.}~\bibnamefont {Xu}},\
  }\bibfield  {title} {\bibinfo {title} {Thermodynamics of noncommutative
  high\textendash{}dimensional ads black holes with non\textendash{}gaussian
  smeared matter distributions},\ }\href
  {https://doi.org/10.1140/epjc/s10052-016-4073-1} {\bibfield  {journal}
  {\bibinfo  {journal} {Eur. Phys. J. C}\ }\textbf {\bibinfo {volume} {76}},\
  \bibinfo {pages} {217} (\bibinfo {year} {2016})}\BibitemShut {NoStop}%
\bibitem [{\citenamefont {Yan}\ \emph {et~al.}(2020)\citenamefont {Yan},
  \citenamefont {Wu},\ and\ \citenamefont {Guo}}]{bib21ad1}%
  \BibitemOpen
  \bibfield  {author} {\bibinfo {author} {\bibfnamefont {Z.}~\bibnamefont
  {Yan}}, \bibinfo {author} {\bibfnamefont {C.}~\bibnamefont {Wu}},\ and\
  \bibinfo {author} {\bibfnamefont {W.}~\bibnamefont {Guo}},\ }\bibfield
  {title} {\bibinfo {title} {Scalar field quasinormal modes of noncommutative
  high dimensional schwarzschild-tangherlini black hole spacetime with smeared
  matter sources},\ }\href {https://doi.org/10.1016/j.nuclphysb.2020.115217}
  {\bibfield  {journal} {\bibinfo  {journal} {Nucl. Phys. B}\ }\textbf
  {\bibinfo {volume} {961}},\ \bibinfo {pages} {115217} (\bibinfo {year}
  {2020})}\BibitemShut {NoStop}%
\bibitem [{\citenamefont {Yan}\ \emph {et~al.}(2021)\citenamefont {Yan},
  \citenamefont {Wu},\ and\ \citenamefont {Guo}}]{bib21ad2}%
  \BibitemOpen
  \bibfield  {author} {\bibinfo {author} {\bibfnamefont {Z.}~\bibnamefont
  {Yan}}, \bibinfo {author} {\bibfnamefont {C.}~\bibnamefont {Wu}},\ and\
  \bibinfo {author} {\bibfnamefont {W.}~\bibnamefont {Guo}},\ }\bibfield
  {title} {\bibinfo {title} {Quasinormal modes of scalar field coupled to
  einstein's tensor in the non-commutative geometry inspired black hole},\
  }\href {https://doi.org/10.1016/j.nuclphysb.2021.115595} {\bibfield
  {journal} {\bibinfo  {journal} {Nucl. Phys. B}\ }\textbf {\bibinfo {volume}
  {973}},\ \bibinfo {pages} {115595} (\bibinfo {year} {2021})}\BibitemShut
  {NoStop}%
\bibitem [{\citenamefont {Bronnikov}\ \emph {et~al.}(2012)\citenamefont
  {Bronnikov}, \citenamefont {Konoplya},\ and\ \citenamefont
  {Zhidenko}}]{bib21ad3}%
  \BibitemOpen
  \bibfield  {author} {\bibinfo {author} {\bibfnamefont {K.~A.}\ \bibnamefont
  {Bronnikov}}, \bibinfo {author} {\bibfnamefont {R.~A.}\ \bibnamefont
  {Konoplya}},\ and\ \bibinfo {author} {\bibfnamefont {A.}~\bibnamefont
  {Zhidenko}},\ }\bibfield  {title} {\bibinfo {title} {Instabilities of
  wormholes and regular black holes supported by a phantom scalar field},\
  }\href {https://doi.org/10.1103/PhysRevD.86.024028} {\bibfield  {journal}
  {\bibinfo  {journal} {Phys. Rev. D}\ }\textbf {\bibinfo {volume} {86}},\
  \bibinfo {pages} {024028} (\bibinfo {year} {2012})}\BibitemShut {NoStop}%
\bibitem [{\citenamefont {Konoplya}\ \emph {et~al.}(2022)\citenamefont
  {Konoplya}, \citenamefont {Zinhailo}, \citenamefont {Kunz}, \citenamefont
  {Stuchlik},\ and\ \citenamefont {Zhidenko}}]{bib21ad4}%
  \BibitemOpen
  \bibfield  {author} {\bibinfo {author} {\bibfnamefont {R.~A.}\ \bibnamefont
  {Konoplya}}, \bibinfo {author} {\bibfnamefont {A.~F.}\ \bibnamefont
  {Zinhailo}}, \bibinfo {author} {\bibfnamefont {J.}~\bibnamefont {Kunz}},
  \bibinfo {author} {\bibfnamefont {Z.}~\bibnamefont {Stuchlik}},\ and\
  \bibinfo {author} {\bibfnamefont {A.}~\bibnamefont {Zhidenko}},\ }\bibfield
  {title} {\bibinfo {title} {Quasinormal ringing of regular black holes in
  asymptotically safe gravity: the importance of overtones},\ }\href
  {https://doi.org/10.1088/1475-7516/2022/10/091} {\bibfield  {journal}
  {\bibinfo  {journal} {JCAP}\ }\textbf {\bibinfo {volume} {10}},\ \bibinfo
  {pages} {091}}\BibitemShut {NoStop}%
\bibitem [{\citenamefont {Futterman}\ \emph {et~al.}(1988)\citenamefont
  {Futterman}, \citenamefont {Handler},\ and\ \citenamefont {Matzner}}]{bib22}%
  \BibitemOpen
  \bibfield  {author} {\bibinfo {author} {\bibfnamefont {J.}~\bibnamefont
  {Futterman}}, \bibinfo {author} {\bibfnamefont {F.}~\bibnamefont {Handler}},\
  and\ \bibinfo {author} {\bibfnamefont {R.~A.}\ \bibnamefont {Matzner}},\
  }\href@noop {} {\emph {\bibinfo {title} {Scattering from black holes}}}\
  (\bibinfo  {publisher} {Cambridge University Press},\ \bibinfo {address}
  {Cambridge; New York},\ \bibinfo {year} {1988})\BibitemShut {NoStop}%
\bibitem [{\citenamefont {Glampedakis}\ and\ \citenamefont
  {Andersson}(2001)}]{bib23}%
  \BibitemOpen
  \bibfield  {author} {\bibinfo {author} {\bibfnamefont {K.}~\bibnamefont
  {Glampedakis}}\ and\ \bibinfo {author} {\bibfnamefont {N.}~\bibnamefont
  {Andersson}},\ }\bibfield  {title} {\bibinfo {title} {Scattering of scalar
  waves by rotating black holes},\ }\href
  {https://doi.org/10.1088/0264-9381/18/10/309} {\bibfield  {journal} {\bibinfo
   {journal} {Class.Quant.Grav.}\ }\textbf {\bibinfo {volume} {18}},\ \bibinfo
  {pages} {1939} (\bibinfo {year} {2001})}\BibitemShut {NoStop}%
\bibitem [{\citenamefont {Doran}\ \emph
  {et~al.}(2005{\natexlab{a}})\citenamefont {Doran}, \citenamefont {Lasenby},
  \citenamefont {Dolan},\ and\ \citenamefont {Hinder}}]{bib24}%
  \BibitemOpen
  \bibfield  {author} {\bibinfo {author} {\bibfnamefont {C.}~\bibnamefont
  {Doran}}, \bibinfo {author} {\bibfnamefont {A.}~\bibnamefont {Lasenby}},
  \bibinfo {author} {\bibfnamefont {S.}~\bibnamefont {Dolan}},\ and\ \bibinfo
  {author} {\bibfnamefont {I.}~\bibnamefont {Hinder}},\ }\bibfield  {title}
  {\bibinfo {title} {Fermion absorption cross section of a schwarzschild black
  hole},\ }\href {https://doi.org/10.1103/PhysRevD.71.124020} {\bibfield
  {journal} {\bibinfo  {journal} {Phys. Rev. D}\ }\textbf {\bibinfo {volume}
  {71}},\ \bibinfo {pages} {124020} (\bibinfo {year}
  {2005}{\natexlab{a}})}\BibitemShut {NoStop}%
\bibitem [{\citenamefont {Dolan}\ \emph
  {et~al.}(2006{\natexlab{a}})\citenamefont {Dolan}, \citenamefont {Doran},\
  and\ \citenamefont {Lasenby}}]{bib25}%
  \BibitemOpen
  \bibfield  {author} {\bibinfo {author} {\bibfnamefont {S.}~\bibnamefont
  {Dolan}}, \bibinfo {author} {\bibfnamefont {C.}~\bibnamefont {Doran}},\ and\
  \bibinfo {author} {\bibfnamefont {A.}~\bibnamefont {Lasenby}},\ }\bibfield
  {title} {\bibinfo {title} {Fermion scattering by a schwarzschild black
  hole},\ }\href {https://doi.org/10.1103/PhysRevD.74.064005} {\bibfield
  {journal} {\bibinfo  {journal} {Phys. Rev. D}\ }\textbf {\bibinfo {volume}
  {74}},\ \bibinfo {pages} {064005} (\bibinfo {year}
  {2006}{\natexlab{a}})}\BibitemShut {NoStop}%
\bibitem [{\citenamefont {Crispino}\ \emph
  {et~al.}(2007{\natexlab{a}})\citenamefont {Crispino}, \citenamefont
  {Oliveira}, \citenamefont {Higuchi},\ and\ \citenamefont {Matsas}}]{bib26}%
  \BibitemOpen
  \bibfield  {author} {\bibinfo {author} {\bibfnamefont {L.~C.~B.}\
  \bibnamefont {Crispino}}, \bibinfo {author} {\bibfnamefont {E.~S.}\
  \bibnamefont {Oliveira}}, \bibinfo {author} {\bibfnamefont {A.}~\bibnamefont
  {Higuchi}},\ and\ \bibinfo {author} {\bibfnamefont {G.~E.~A.}\ \bibnamefont
  {Matsas}},\ }\bibfield  {title} {\bibinfo {title} {Absorption cross section
  of electromagnetic waves for schwarzschild black holes},\ }\href
  {https://doi.org/10.1103/PhysRevD.75.104012} {\bibfield  {journal} {\bibinfo
  {journal} {Phys. Rev. D}\ }\textbf {\bibinfo {volume} {75}},\ \bibinfo
  {pages} {104012} (\bibinfo {year} {2007}{\natexlab{a}})}\BibitemShut
  {NoStop}%
\bibitem [{\citenamefont {Dolan}(2008)}]{bib27}%
  \BibitemOpen
  \bibfield  {author} {\bibinfo {author} {\bibfnamefont {S.~R.}\ \bibnamefont
  {Dolan}},\ }\bibfield  {title} {\bibinfo {title} {Scattering and absorption
  of gravitational plane waves by rotating black holes},\ }\href
  {https://doi.org/10.1088/0264-9381/25/23/235002} {\bibfield  {journal}
  {\bibinfo  {journal} {Class.Quant.Grav.}\ }\textbf {\bibinfo {volume} {25}},\
  \bibinfo {pages} {235002} (\bibinfo {year} {2008})}\BibitemShut {NoStop}%
\bibitem [{\citenamefont {Matzner}(1968)}]{bib28}%
  \BibitemOpen
  \bibfield  {author} {\bibinfo {author} {\bibfnamefont {R.~A.}\ \bibnamefont
  {Matzner}},\ }\bibfield  {title} {\bibinfo {title} {Scattering of massless
  scalar waves by a schwarzschild ``singularity''},\ }\href
  {https://doi.org/10.1063/1.1664470} {\bibfield  {journal} {\bibinfo
  {journal} {J. MATH. PHYS.}\ }\textbf {\bibinfo {volume} {9}},\ \bibinfo
  {pages} {163} (\bibinfo {year} {1968})}\BibitemShut {NoStop}%
\bibitem [{\citenamefont {Mashhoon}(1973)}]{bib29}%
  \BibitemOpen
  \bibfield  {author} {\bibinfo {author} {\bibfnamefont {B.}~\bibnamefont
  {Mashhoon}},\ }\bibfield  {title} {\bibinfo {title} {Scattering of
  electromagnetic radiation from a black hole},\ }\href
  {https://doi.org/10.1103/PhysRevD.7.2807} {\bibfield  {journal} {\bibinfo
  {journal} {Phys. Rev. D}\ }\textbf {\bibinfo {volume} {7}},\ \bibinfo {pages}
  {2807} (\bibinfo {year} {1973})}\BibitemShut {NoStop}%
\bibitem [{\citenamefont {Starobinski\v{\i}}(1973)}]{bib30}%
  \BibitemOpen
  \bibfield  {author} {\bibinfo {author} {\bibfnamefont {A.~A.}\ \bibnamefont
  {Starobinski\v{\i}}},\ }\bibfield  {title} {\bibinfo {title} {Amplification
  of waves reflected from a rotating ''black hole''},\ }\href@noop {}
  {\bibfield  {journal} {\bibinfo  {journal} {Zh. Eksp. Teor. Fiz}\ }\textbf
  {\bibinfo {volume} {7}},\ \bibinfo {pages} {28} (\bibinfo {year}
  {1973})}\BibitemShut {NoStop}%
\bibitem [{\citenamefont {Starobinski\v{\i}}\ and\ \citenamefont
  {Churilov}(1973)}]{bib31}%
  \BibitemOpen
  \bibfield  {author} {\bibinfo {author} {\bibfnamefont {A.~A.}\ \bibnamefont
  {Starobinski\v{\i}}}\ and\ \bibinfo {author} {\bibfnamefont {S.~M.}\
  \bibnamefont {Churilov}},\ }\bibfield  {title} {\bibinfo {title}
  {Amplification of electromagnetic and gravitational waves scattered by a
  rotating “black hole”},\ }\href@noop {} {\bibfield  {journal} {\bibinfo
  {journal} {Zh. Eksp. Teor. Fiz}\ }\textbf {\bibinfo {volume} {65}},\ \bibinfo
  {pages} {3} (\bibinfo {year} {1973})}\BibitemShut {NoStop}%
\bibitem [{\citenamefont {Crispino}\ \emph {et~al.}(2001)\citenamefont
  {Crispino}, \citenamefont {Higuchi},\ and\ \citenamefont {Matsas}}]{bib32}%
  \BibitemOpen
  \bibfield  {author} {\bibinfo {author} {\bibfnamefont {L.~C.~B.}\
  \bibnamefont {Crispino}}, \bibinfo {author} {\bibfnamefont {A.}~\bibnamefont
  {Higuchi}},\ and\ \bibinfo {author} {\bibfnamefont {G.~E.~A.}\ \bibnamefont
  {Matsas}},\ }\bibfield  {title} {\bibinfo {title} {Quantization of the
  electromagnetic field outside static black holes and its application to
  low\textendash{}energy phenomena},\ }\href
  {https://doi.org/10.1103/PhysRevD.63.124008} {\bibfield  {journal} {\bibinfo
  {journal} {Phys. Rev. D}\ }\textbf {\bibinfo {volume} {63}},\ \bibinfo
  {pages} {124008} (\bibinfo {year} {2001})}\BibitemShut {NoStop}%
\bibitem [{\citenamefont {Ju\textendash{}Hua}\ and\ \citenamefont
  {Yong\textendash{}Jiu}(2010)}]{bib33}%
  \BibitemOpen
  \bibfield  {author} {\bibinfo {author} {\bibfnamefont {C.}~\bibnamefont
  {Ju\textendash{}Hua}}\ and\ \bibinfo {author} {\bibfnamefont
  {W.}~\bibnamefont {Yong\textendash{}Jiu}},\ }\bibfield  {title} {\bibinfo
  {title} {Quasinormal modes of the scalar field in
  five\textendash{}dimensional lovelock black hole spacetime},\ }\href
  {https://doi.org/10.1088/1674-1056/19/6/060401} {\bibfield  {journal}
  {\bibinfo  {journal} {Chin. Phys. B}\ }\textbf {\bibinfo {volume} {19}},\
  \bibinfo {pages} {060401} (\bibinfo {year} {2010})}\BibitemShut {NoStop}%
\bibitem [{\citenamefont {Kobayashi}\ and\ \citenamefont
  {Tomimatsu}(2010)}]{bib34}%
  \BibitemOpen
  \bibfield  {author} {\bibinfo {author} {\bibfnamefont {T.}~\bibnamefont
  {Kobayashi}}\ and\ \bibinfo {author} {\bibfnamefont {A.}~\bibnamefont
  {Tomimatsu}},\ }\bibfield  {title} {\bibinfo {title} {Superradiant scattering
  of electromagnetic waves emitted from disk around kerr black holes},\ }\href
  {https://doi.org/10.1103/PhysRevD.82.084026} {\bibfield  {journal} {\bibinfo
  {journal} {Phys. Rev. D}\ }\textbf {\bibinfo {volume} {82}},\ \bibinfo
  {pages} {084026} (\bibinfo {year} {2010})}\BibitemShut {NoStop}%
\bibitem [{\citenamefont {Chen}\ \emph {et~al.}(2011)\citenamefont {Chen},
  \citenamefont {Liao},\ and\ \citenamefont {Wang}}]{bib35}%
  \BibitemOpen
  \bibfield  {author} {\bibinfo {author} {\bibfnamefont {J.}~\bibnamefont
  {Chen}}, \bibinfo {author} {\bibfnamefont {H.}~\bibnamefont {Liao}},\ and\
  \bibinfo {author} {\bibfnamefont {Y.}~\bibnamefont {Wang}},\ }\bibfield
  {title} {\bibinfo {title} {Absorption of massless scalar wave by
  high\textendash{}dimensional lovelock black hole},\ }\href
  {https://doi.org/10.1016/j.physletb.2011.09.091} {\bibfield  {journal}
  {\bibinfo  {journal} {Phys. Lett. B}\ }\textbf {\bibinfo {volume} {705}},\
  \bibinfo {pages} {124} (\bibinfo {year} {2011})}\BibitemShut {NoStop}%
\bibitem [{\citenamefont {Liao}\ \emph {et~al.}(2012)\citenamefont {Liao},
  \citenamefont {Chen},\ and\ \citenamefont {Wang}}]{bib36}%
  \BibitemOpen
  \bibfield  {author} {\bibinfo {author} {\bibfnamefont {H.}~\bibnamefont
  {Liao}}, \bibinfo {author} {\bibfnamefont {J.}~\bibnamefont {Chen}},\ and\
  \bibinfo {author} {\bibfnamefont {Y.}~\bibnamefont {Wang}},\ }\bibfield
  {title} {\bibinfo {title} {Scattering of scalar wave from black hole in
  horava\textendash{}lifshitz gravity},\ }\href
  {https://doi.org/10.1142/S0218271812500459} {\bibfield  {journal} {\bibinfo
  {journal} {Int. J. Mod. Phys. D}\ }\textbf {\bibinfo {volume} {21}},\
  \bibinfo {pages} {1250045} (\bibinfo {year} {2012})}\BibitemShut {NoStop}%
\bibitem [{\citenamefont {Frolov}\ and\ \citenamefont {Shoom}(2012)}]{bib37}%
  \BibitemOpen
  \bibfield  {author} {\bibinfo {author} {\bibfnamefont {V.~P.}\ \bibnamefont
  {Frolov}}\ and\ \bibinfo {author} {\bibfnamefont {A.~A.}\ \bibnamefont
  {Shoom}},\ }\bibfield  {title} {\bibinfo {title} {Scattering of circularly
  polarized light by a rotating black hole},\ }\href
  {https://doi.org/10.1103/PhysRevD.86.024010} {\bibfield  {journal} {\bibinfo
  {journal} {Phys. Rev. D}\ }\textbf {\bibinfo {volume} {86}},\ \bibinfo
  {pages} {024010} (\bibinfo {year} {2012})}\BibitemShut {NoStop}%
\bibitem [{\citenamefont {S\'{a}nchez}(1977)}]{bib38}%
  \BibitemOpen
  \bibfield  {author} {\bibinfo {author} {\bibfnamefont {N.}~\bibnamefont
  {S\'{a}nchez}},\ }\bibfield  {title} {\bibinfo {title} {Wave scattering
  theory and the absorption problem for a black hole},\ }\href
  {https://doi.org/10.1103/PhysRevD.16.937} {\bibfield  {journal} {\bibinfo
  {journal} {Phys. Rev. D}\ }\textbf {\bibinfo {volume} {16}},\ \bibinfo
  {pages} {937} (\bibinfo {year} {1977})}\BibitemShut {NoStop}%
\bibitem [{\citenamefont {S\'anchez}(1978)}]{bib39}%
  \BibitemOpen
  \bibfield  {author} {\bibinfo {author} {\bibfnamefont {N.}~\bibnamefont
  {S\'anchez}},\ }\bibfield  {title} {\bibinfo {title} {Absorption and emission
  spectra of a schwarzschild black hole},\ }\href
  {https://doi.org/10.1103/PhysRevD.18.1030} {\bibfield  {journal} {\bibinfo
  {journal} {Phys. Rev. D}\ }\textbf {\bibinfo {volume} {18}},\ \bibinfo
  {pages} {1030} (\bibinfo {year} {1978})}\BibitemShut {NoStop}%
\bibitem [{\citenamefont {S\'{a}nchez}(1978)}]{bib40}%
  \BibitemOpen
  \bibfield  {author} {\bibinfo {author} {\bibfnamefont {N.}~\bibnamefont
  {S\'{a}nchez}},\ }\bibfield  {title} {\bibinfo {title} {Elastic scattering of
  waves by a black hole},\ }\href {https://doi.org/10.1103/PhysRevD.18.1798}
  {\bibfield  {journal} {\bibinfo  {journal} {Phys. Rev. D}\ }\textbf {\bibinfo
  {volume} {18}},\ \bibinfo {pages} {1798} (\bibinfo {year}
  {1978})}\BibitemShut {NoStop}%
\bibitem [{\citenamefont {Jung}\ \emph {et~al.}(2004)\citenamefont {Jung},
  \citenamefont {Kim},\ and\ \citenamefont {Park}}]{bib41}%
  \BibitemOpen
  \bibfield  {author} {\bibinfo {author} {\bibfnamefont {E.}~\bibnamefont
  {Jung}}, \bibinfo {author} {\bibfnamefont {S.}~\bibnamefont {Kim}},\ and\
  \bibinfo {author} {\bibfnamefont {D.~K.}\ \bibnamefont {Park}},\ }\bibfield
  {title} {\bibinfo {title} {Proof of universality for the absorption of
  massive scalar by the higher\textendash{}dimensional
  reissner\textendash{}nordstr{\"o}m black holes},\ }\href
  {https://doi.org/10.1016/j.physletb.2004.09.067} {\bibfield  {journal}
  {\bibinfo  {journal} {Phys. Lett. B}\ }\textbf {\bibinfo {volume} {602}},\
  \bibinfo {pages} {105} (\bibinfo {year} {2004})}\BibitemShut {NoStop}%
\bibitem [{\citenamefont {Doran}\ \emph
  {et~al.}(2005{\natexlab{b}})\citenamefont {Doran}, \citenamefont {Lasenby},
  \citenamefont {Dolan},\ and\ \citenamefont {Hinder}}]{bib42}%
  \BibitemOpen
  \bibfield  {author} {\bibinfo {author} {\bibfnamefont {C.}~\bibnamefont
  {Doran}}, \bibinfo {author} {\bibfnamefont {A.}~\bibnamefont {Lasenby}},
  \bibinfo {author} {\bibfnamefont {S.}~\bibnamefont {Dolan}},\ and\ \bibinfo
  {author} {\bibfnamefont {I.}~\bibnamefont {Hinder}},\ }\bibfield  {title}
  {\bibinfo {title} {Fermion absorption cross section of a schwarzschild black
  hole},\ }\href {https://doi.org/10.1103/PhysRevD.71.124020} {\bibfield
  {journal} {\bibinfo  {journal} {Phys. Rev. D}\ }\textbf {\bibinfo {volume}
  {71}},\ \bibinfo {pages} {124020} (\bibinfo {year}
  {2005}{\natexlab{b}})}\BibitemShut {NoStop}%
\bibitem [{\citenamefont {Dolan}\ \emph
  {et~al.}(2006{\natexlab{b}})\citenamefont {Dolan}, \citenamefont {Doran},\
  and\ \citenamefont {Lasenby}}]{bib43}%
  \BibitemOpen
  \bibfield  {author} {\bibinfo {author} {\bibfnamefont {S.}~\bibnamefont
  {Dolan}}, \bibinfo {author} {\bibfnamefont {C.}~\bibnamefont {Doran}},\ and\
  \bibinfo {author} {\bibfnamefont {A.}~\bibnamefont {Lasenby}},\ }\bibfield
  {title} {\bibinfo {title} {Fermion scattering by a schwarzschild black
  hole},\ }\href {https://doi.org/10.1103/PhysRevD.74.064005} {\bibfield
  {journal} {\bibinfo  {journal} {Phys. Rev. D}\ }\textbf {\bibinfo {volume}
  {74}},\ \bibinfo {pages} {064005} (\bibinfo {year}
  {2006}{\natexlab{b}})}\BibitemShut {NoStop}%
\bibitem [{\citenamefont {Castineiras}\ \emph {et~al.}(2007)\citenamefont
  {Castineiras}, \citenamefont {Crispino},\ and\ \citenamefont
  {MeiraFilho}}]{bib44}%
  \BibitemOpen
  \bibfield  {author} {\bibinfo {author} {\bibfnamefont {J.}~\bibnamefont
  {Castineiras}}, \bibinfo {author} {\bibfnamefont {L.~C.~B.}\ \bibnamefont
  {Crispino}},\ and\ \bibinfo {author} {\bibfnamefont {D.~P.}\ \bibnamefont
  {MeiraFilho}},\ }\bibfield  {title} {\bibinfo {title} {Source coupled to the
  massive scalar field orbiting a stellar object},\ }\href
  {https://doi.org/10.1103/PhysRevD.75.024012} {\bibfield  {journal} {\bibinfo
  {journal} {Phys. Rev. D}\ }\textbf {\bibinfo {volume} {75}},\ \bibinfo
  {pages} {024012} (\bibinfo {year} {2007})}\BibitemShut {NoStop}%
\bibitem [{\citenamefont {Benone}\ \emph {et~al.}(2017)\citenamefont {Benone},
  \citenamefont {de~Oliveira}, \citenamefont {Dolan},\ and\ \citenamefont
  {Crispino}}]{bib45}%
  \BibitemOpen
  \bibfield  {author} {\bibinfo {author} {\bibfnamefont {C.~L.}\ \bibnamefont
  {Benone}}, \bibinfo {author} {\bibfnamefont {E.~S.}\ \bibnamefont
  {de~Oliveira}}, \bibinfo {author} {\bibfnamefont {S.~R.}\ \bibnamefont
  {Dolan}},\ and\ \bibinfo {author} {\bibfnamefont {L.~C.~B.}\ \bibnamefont
  {Crispino}},\ }\bibfield  {title} {\bibinfo {title} {Addendum to ``absorption
  of a massive scalar field by a charged black hole''},\ }\href
  {https://doi.org/10.1103/PhysRevD.95.044035} {\bibfield  {journal} {\bibinfo
  {journal} {Phys. Rev. D}\ }\textbf {\bibinfo {volume} {95}},\ \bibinfo
  {pages} {044035} (\bibinfo {year} {2017})}\BibitemShut {NoStop}%
\bibitem [{\citenamefont {Chen}\ \emph {et~al.}(2013)\citenamefont {Chen},
  \citenamefont {Liao}, \citenamefont {Wang},\ and\ \citenamefont
  {Chen}}]{bib46}%
  \BibitemOpen
  \bibfield  {author} {\bibinfo {author} {\bibfnamefont {J.}~\bibnamefont
  {Chen}}, \bibinfo {author} {\bibfnamefont {H.}~\bibnamefont {Liao}}, \bibinfo
  {author} {\bibfnamefont {Y.}~\bibnamefont {Wang}},\ and\ \bibinfo {author}
  {\bibfnamefont {T.}~\bibnamefont {Chen}},\ }\bibfield  {title} {\bibinfo
  {title} {Absorption and scattering of scalar wave from schwarzschild black
  hole surrounded by magnetic field},\ }\bibfield  {journal} {\bibinfo
  {journal} {The European Physical Journal C}\ }\textbf {\bibinfo {volume}
  {73}},\ \href {https://doi.org/10.1140/epjc/s10052-013-2395-9}
  {10.1140/epjc/s10052-013-2395-9} (\bibinfo {year} {2013})\BibitemShut
  {NoStop}%
\bibitem [{\citenamefont {Crispino}\ \emph {et~al.}(2009)\citenamefont
  {Crispino}, \citenamefont {Dolan},\ and\ \citenamefont {Oliveira}}]{bib47}%
  \BibitemOpen
  \bibfield  {author} {\bibinfo {author} {\bibfnamefont {L.~C.~B.}\
  \bibnamefont {Crispino}}, \bibinfo {author} {\bibfnamefont {S.~R.}\
  \bibnamefont {Dolan}},\ and\ \bibinfo {author} {\bibfnamefont {E.~S.}\
  \bibnamefont {Oliveira}},\ }\bibfield  {title} {\bibinfo {title} {Scattering
  of massless scalar waves by reissner\textendash{}nordstr{\"o}m black holes},\
  }\href {https://doi.org/10.1103/PhysRevD.79.064022} {\bibfield  {journal}
  {\bibinfo  {journal} {Phys. Rev. D}\ }\textbf {\bibinfo {volume} {79}},\
  \bibinfo {pages} {064022} (\bibinfo {year} {2009})}\BibitemShut {NoStop}%
\bibitem [{\citenamefont {Huang}\ \emph {et~al.}(2014)\citenamefont {Huang},
  \citenamefont {Liao}, \citenamefont {Chen},\ and\ \citenamefont
  {Wang}}]{bib48}%
  \BibitemOpen
  \bibfield  {author} {\bibinfo {author} {\bibfnamefont {H.}~\bibnamefont
  {Huang}}, \bibinfo {author} {\bibfnamefont {P.}~\bibnamefont {Liao}},
  \bibinfo {author} {\bibfnamefont {J.}~\bibnamefont {Chen}},\ and\ \bibinfo
  {author} {\bibfnamefont {Y.}~\bibnamefont {Wang}},\ }\bibfield  {title}
  {\bibinfo {title} {Absorption and scattering cross section of regular black
  holes},\ }\bibfield  {journal} {\bibinfo  {journal} {J. Grav.}\ }\textbf
  {\bibinfo {volume} {2014}},\ \href {https://doi.org/10.1155/2014/231727}
  {10.1155/2014/231727} (\bibinfo {year} {2014})\BibitemShut {NoStop}%
\bibitem [{\citenamefont {Anacleto}\ \emph {et~al.}(2020)\citenamefont
  {Anacleto}, \citenamefont {Brito}, \citenamefont {Campos},\ and\
  \citenamefont {Passos}}]{bib49}%
  \BibitemOpen
  \bibfield  {author} {\bibinfo {author} {\bibfnamefont {M.}~\bibnamefont
  {Anacleto}}, \bibinfo {author} {\bibfnamefont {F.}~\bibnamefont {Brito}},
  \bibinfo {author} {\bibfnamefont {J.}~\bibnamefont {Campos}},\ and\ \bibinfo
  {author} {\bibfnamefont {E.}~\bibnamefont {Passos}},\ }\bibfield  {title}
  {\bibinfo {title} {Absorption and scattering of a noncommutative black
  hole},\ }\href {https://doi.org/10.1016/j.physletb.2020.135334} {\bibfield
  {journal} {\bibinfo  {journal} {Phys. Lett. B}\ }\textbf {\bibinfo {volume}
  {803}},\ \bibinfo {pages} {135334} (\bibinfo {year} {2020})}\BibitemShut
  {NoStop}%
\bibitem [{\citenamefont {Anacleto}\ \emph {et~al.}(2015)\citenamefont
  {Anacleto}, \citenamefont {Brito},\ and\ \citenamefont {Passos}}]{bib50}%
  \BibitemOpen
  \bibfield  {author} {\bibinfo {author} {\bibfnamefont {M.~A.}\ \bibnamefont
  {Anacleto}}, \bibinfo {author} {\bibfnamefont {F.~A.}\ \bibnamefont
  {Brito}},\ and\ \bibinfo {author} {\bibfnamefont {E.}~\bibnamefont
  {Passos}},\ }\bibfield  {title} {\bibinfo {title} {Gravitational
  aharonov\textendash{}bohm effect due to noncommutative btz black hole},\
  }\href {https://doi.org/10.1016/j.physletb.2015.02.056} {\bibfield  {journal}
  {\bibinfo  {journal} {Phys. Lett. B}\ }\textbf {\bibinfo {volume} {743}},\
  \bibinfo {pages} {184} (\bibinfo {year} {2015})}\BibitemShut {NoStop}%
\bibitem [{\citenamefont {Moura}(2013)}]{bib51}%
  \BibitemOpen
  \bibfield  {author} {\bibinfo {author} {\bibfnamefont {F.}~\bibnamefont
  {Moura}},\ }\bibfield  {title} {\bibinfo {title} {Scattering of spherically
  symmetric $d$\textendash{}dimensional $\alpha'$\textendash{}corrected black
  holes in string theory},\ }\href {https://doi.org/10.1007/JHEP09(2013)038}
  {\bibfield  {journal} {\bibinfo  {journal} {JHEP}\ }\textbf {\bibinfo
  {volume} {09}},\ \bibinfo {pages} {038}}\BibitemShut {NoStop}%
\bibitem [{\citenamefont {Nicolini}\ \emph {et~al.}(2013)\citenamefont
  {Nicolini}, \citenamefont {Orlandi},\ and\ \citenamefont
  {Spallucci}}]{bib52}%
  \BibitemOpen
  \bibfield  {author} {\bibinfo {author} {\bibfnamefont {P.}~\bibnamefont
  {Nicolini}}, \bibinfo {author} {\bibfnamefont {A.}~\bibnamefont {Orlandi}},\
  and\ \bibinfo {author} {\bibfnamefont {E.}~\bibnamefont {Spallucci}},\
  }\bibfield  {title} {\bibinfo {title} {The final stage of gravitationally
  collapsed thick matter layers},\ }\href {https://doi.org/10.1155/2013/812084}
  {\bibfield  {journal} {\bibinfo  {journal} {Adv. High Energy Phys.}\ }\textbf
  {\bibinfo {volume} {2013}},\ \bibinfo {pages} {812084} (\bibinfo {year}
  {2013})}\BibitemShut {NoStop}%
\bibitem [{\citenamefont {Miao}\ and\ \citenamefont {Wu}(2017)}]{bib53}%
  \BibitemOpen
  \bibfield  {author} {\bibinfo {author} {\bibfnamefont {Y.}~\bibnamefont
  {Miao}}\ and\ \bibinfo {author} {\bibfnamefont {Y.}~\bibnamefont {Wu}},\
  }\bibfield  {title} {\bibinfo {title} {Thermodynamics of the
  schwarzschild\textendash{}ads black hole with a minimal length},\ }\href
  {https://doi.org/10.1155/2017/1095217} {\bibfield  {journal} {\bibinfo
  {journal} {Adv. High Energy Phys.}\ }\textbf {\bibinfo {volume} {2017}},\
  \bibinfo {pages} {1095217} (\bibinfo {year} {2017})}\BibitemShut {NoStop}%
\bibitem [{\citenamefont {Miao}\ and\ \citenamefont {Xu}(2017)}]{bib54}%
  \BibitemOpen
  \bibfield  {author} {\bibinfo {author} {\bibfnamefont {Y.}~\bibnamefont
  {Miao}}\ and\ \bibinfo {author} {\bibfnamefont {Z.}~\bibnamefont {Xu}},\
  }\bibfield  {title} {\bibinfo {title} {Phase transition and entropy
  inequality of noncommutative black holes in a new extended phase space},\
  }\href {https://doi.org/10.1088/1475-7516/2017/03/046} {\bibfield  {journal}
  {\bibinfo  {journal} {JCAP}\ }\textbf {\bibinfo {volume} {2017}}\bibinfo
  {number} { (03)},\ \bibinfo {pages} {046}}\BibitemShut {NoStop}%
\bibitem [{\citenamefont {Park}(2009)}]{bib55}%
  \BibitemOpen
\bibfield  {number} {  }\bibfield  {author} {\bibinfo {author} {\bibfnamefont
  {M.}~\bibnamefont {Park}},\ }\bibfield  {title} {\bibinfo {title} {Smeared
  hair and black holes in three\textendash{}dimensional de sitter spacetime},\
  }\href {https://doi.org/10.1103/PhysRevD.80.084026} {\bibfield  {journal}
  {\bibinfo  {journal} {Phys. Rev. D}\ }\textbf {\bibinfo {volume} {80}},\
  \bibinfo {pages} {084026} (\bibinfo {year} {2009})}\BibitemShut {NoStop}%
\bibitem [{\citenamefont {Wu}\ and\ \citenamefont {Miao}(2022)}]{bib56}%
  \BibitemOpen
  \bibfield  {author} {\bibinfo {author} {\bibfnamefont {Y.}~\bibnamefont
  {Wu}}\ and\ \bibinfo {author} {\bibfnamefont {Y.}~\bibnamefont {Miao}},\
  }\bibfield  {title} {\bibinfo {title} {Higher\textendash{}dimensional regular
  reissner\textendash{}nordstr{\"o}m black holes associated with linear
  electrodynamics},\ }\href {https://doi.org/10.3390/universe8010043}
  {\bibfield  {journal} {\bibinfo  {journal} {Universe}\ }\textbf {\bibinfo
  {volume} {8}},\ \bibinfo {pages} {43} (\bibinfo {year} {2022})}\BibitemShut
  {NoStop}%
\bibitem [{\citenamefont {Cardoso}\ \emph {et~al.}(2003)\citenamefont
  {Cardoso}, \citenamefont {Dias},\ and\ \citenamefont {Lemos}}]{bib57}%
  \BibitemOpen
  \bibfield  {author} {\bibinfo {author} {\bibfnamefont {V.}~\bibnamefont
  {Cardoso}}, \bibinfo {author} {\bibfnamefont {O.~J.~C.}\ \bibnamefont
  {Dias}},\ and\ \bibinfo {author} {\bibfnamefont {J.~P.~S.}\ \bibnamefont
  {Lemos}},\ }\bibfield  {title} {\bibinfo {title} {Gravitational radiation in
  $d$\textendash{}dimensional space\textendash{}times},\ }\href
  {https://doi.org/10.1103/PhysRevD.67.064026} {\bibfield  {journal} {\bibinfo
  {journal} {Phys. Rev. D}\ }\textbf {\bibinfo {volume} {67}},\ \bibinfo
  {pages} {064026} (\bibinfo {year} {2003})}\BibitemShut {NoStop}%
\bibitem [{\citenamefont {Konoplya}(2003)}]{bib58}%
  \BibitemOpen
  \bibfield  {author} {\bibinfo {author} {\bibfnamefont {R.~A.}\ \bibnamefont
  {Konoplya}},\ }\bibfield  {title} {\bibinfo {title} {Quasinormal behavior of
  the $d$\textendash{}dimensional schwarzschild black hole and higher order wkb
  approach},\ }\href {https://doi.org/10.1103/PhysRevD.68.024018} {\bibfield
  {journal} {\bibinfo  {journal} {Phys. Rev. D}\ }\textbf {\bibinfo {volume}
  {68}},\ \bibinfo {pages} {024018} (\bibinfo {year} {2003})}\BibitemShut
  {NoStop}%
\bibitem [{\citenamefont {Tangherlini}(1963)}]{bib59}%
  \BibitemOpen
  \bibfield  {author} {\bibinfo {author} {\bibfnamefont {F.~R.}\ \bibnamefont
  {Tangherlini}},\ }\bibfield  {title} {\bibinfo {title} {Schwarzschild field
  in n dimensions and the dimensionality of space problem},\ }\href
  {https://doi.org/10.1007/BF02784569} {\bibfield  {journal} {\bibinfo
  {journal} {Nuovo Cim.}\ }\textbf {\bibinfo {volume} {27}},\ \bibinfo {pages}
  {636} (\bibinfo {year} {1963})}\BibitemShut {NoStop}%
\bibitem [{\citenamefont {Natario}\ and\ \citenamefont
  {Schiappa}(2004)}]{bib60}%
  \BibitemOpen
  \bibfield  {author} {\bibinfo {author} {\bibfnamefont {J.}~\bibnamefont
  {Natario}}\ and\ \bibinfo {author} {\bibfnamefont {R.}~\bibnamefont
  {Schiappa}},\ }\bibfield  {title} {\bibinfo {title} {On the classification of
  asymptotic quasinormal frequencies for $d$\textendash{}dimensional black
  holes and quantum gravity},\ }\href
  {https://doi.org/10.4310/ATMP.2004.v8.n6.a4} {\bibfield  {journal} {\bibinfo
  {journal} {Adv. Theor. Math. Phys.}\ }\textbf {\bibinfo {volume} {8}},\
  \bibinfo {pages} {1001} (\bibinfo {year} {2004})}\BibitemShut {NoStop}%
\bibitem [{\citenamefont {Casadio}\ \emph {et~al.}(2014)\citenamefont
  {Casadio}, \citenamefont {Micu},\ and\ \citenamefont {Scardigli}}]{bib61}%
  \BibitemOpen
  \bibfield  {author} {\bibinfo {author} {\bibfnamefont {R.}~\bibnamefont
  {Casadio}}, \bibinfo {author} {\bibfnamefont {O.}~\bibnamefont {Micu}},\ and\
  \bibinfo {author} {\bibfnamefont {F.}~\bibnamefont {Scardigli}},\ }\bibfield
  {title} {\bibinfo {title} {Quantum hoop conjecture: Black hole formation by
  particle collisions},\ }\href
  {https://doi.org/10.1016/j.physletb.2014.03.037} {\bibfield  {journal}
  {\bibinfo  {journal} {Phys. Lett. B}\ }\textbf {\bibinfo {volume} {732}},\
  \bibinfo {pages} {105} (\bibinfo {year} {2014})}\BibitemShut {NoStop}%
\bibitem [{\citenamefont {Thorne}\ and\ \citenamefont {Klauder}(1972)}]{bib62}%
  \BibitemOpen
  \bibfield  {author} {\bibinfo {author} {\bibfnamefont {K.~S.}\ \bibnamefont
  {Thorne}}\ and\ \bibinfo {author} {\bibfnamefont {J.}~\bibnamefont
  {Klauder}},\ }\href@noop {} {\emph {\bibinfo {title} {Magic Without Magic:
  John Archibald Wheeler}}}\ (\bibinfo  {publisher} {Freeman},\ \bibinfo
  {address} {San Francisco},\ \bibinfo {year} {1972})\BibitemShut {NoStop}%
\bibitem [{\citenamefont {Perlick}\ and\ \citenamefont {Tsupko}(2022)}]{bib63}%
  \BibitemOpen
  \bibfield  {author} {\bibinfo {author} {\bibfnamefont {V.}~\bibnamefont
  {Perlick}}\ and\ \bibinfo {author} {\bibfnamefont {O.~Y.}\ \bibnamefont
  {Tsupko}},\ }\bibfield  {title} {\bibinfo {title} {Calculating black hole
  shadows: Review of analytical studies},\ }\href
  {https://doi.org/10.1016/j.physrep.2021.10.004} {\bibfield  {journal}
  {\bibinfo  {journal} {Physics Reports}\ }\textbf {\bibinfo {volume} {947}},\
  \bibinfo {pages} {1} (\bibinfo {year} {2022})}\BibitemShut {NoStop}%
\bibitem [{\citenamefont {Vagnozzi}\ \emph {et~al.}(2022)\citenamefont
  {Vagnozzi} \emph {et~al.}}]{bib68}%
  \BibitemOpen
  \bibfield  {author} {\bibinfo {author} {\bibfnamefont {S.}~\bibnamefont
  {Vagnozzi}} \emph {et~al.},\ }\bibfield  {title} {\bibinfo {title}
  {{Horizon-scale tests of gravity theories and fundamental physics from the
  Event Horizon Telescope image of Sagittarius A$^*$}},\ }\href@noop {} {\
  (\bibinfo {year} {2022})},\ \Eprint {https://arxiv.org/abs/2205.07787}
  {arXiv:2205.07787 [gr-qc]} \BibitemShut {NoStop}%
\bibitem [{\citenamefont {et~al. (Event Horizon~Telescope)}(2022)}]{bib69}%
  \BibitemOpen
  \bibfield  {author} {\bibinfo {author} {\bibfnamefont {K.~A.}\ \bibnamefont
  {et~al. (Event Horizon~Telescope)}},\ }\bibfield  {title} {\bibinfo {title}
  {First sagittarius a* event horizon telescope results. vi. testing the black
  hole metric},\ }\href {https://doi.org/10.3847/2041-8213/ac6756} {\bibfield
  {journal} {\bibinfo  {journal} {The Astrophysical Journal Letters}\ }\textbf
  {\bibinfo {volume} {930}},\ \bibinfo {pages} {L17} (\bibinfo {year}
  {2022})}\BibitemShut {NoStop}%
\bibitem [{\citenamefont {Crispino}\ \emph
  {et~al.}(2007{\natexlab{b}})\citenamefont {Crispino}, \citenamefont
  {Oliveira},\ and\ \citenamefont {Matsas}}]{bib70}%
  \BibitemOpen
  \bibfield  {author} {\bibinfo {author} {\bibfnamefont {L.~C.~B.}\
  \bibnamefont {Crispino}}, \bibinfo {author} {\bibfnamefont {E.~S.}\
  \bibnamefont {Oliveira}},\ and\ \bibinfo {author} {\bibfnamefont {G.~E.~A.}\
  \bibnamefont {Matsas}},\ }\bibfield  {title} {\bibinfo {title} {Absorption
  cross section of canonical acoustic holes},\ }\href
  {https://doi.org/10.1103/PhysRevD.76.107502} {\bibfield  {journal} {\bibinfo
  {journal} {Phys. Rev. D}\ }\textbf {\bibinfo {volume} {76}},\ \bibinfo
  {pages} {107502} (\bibinfo {year} {2007}{\natexlab{b}})}\BibitemShut
  {NoStop}%
\bibitem [{\citenamefont {Dolan}\ \emph {et~al.}(2009)\citenamefont {Dolan},
  \citenamefont {Oliveira},\ and\ \citenamefont {Crispino}}]{bib71}%
  \BibitemOpen
  \bibfield  {author} {\bibinfo {author} {\bibfnamefont {S.~R.}\ \bibnamefont
  {Dolan}}, \bibinfo {author} {\bibfnamefont {E.~S.}\ \bibnamefont
  {Oliveira}},\ and\ \bibinfo {author} {\bibfnamefont {L.~C.~B.}\ \bibnamefont
  {Crispino}},\ }\bibfield  {title} {\bibinfo {title} {Scattering of sound
  waves by a canonical acoustic hole},\ }\href
  {https://doi.org/10.1103/PhysRevD.79.064014} {\bibfield  {journal} {\bibinfo
  {journal} {Phys. Rev. D}\ }\textbf {\bibinfo {volume} {79}},\ \bibinfo
  {pages} {064014} (\bibinfo {year} {2009})}\BibitemShut {NoStop}%
\end{thebibliography}%

\end{document}